\documentstyle[aps,epsf]{revtex}
\def\beq{\begin{equation}}
\def\eeq{\end{equation}}
\def\barr{\begin{eqnarray}}
\def\earr{\end{eqnarray}}
\def\winf{W_{1+\infty}\ }
\def\u1{\widehat{U(1)}}
\def\su2{\widehat{SU(2)}_1}

\def\rr{\rangle\rangle}
\newcommand{\nl}{\nonumber \\}

\newcommand{\ga}{\gamma}

\begin{document}
\preprint{DFF 300/06/98 }

\title{Numerical Study of Hierarchical Quantum Hall
Edge States on the Disk Geometry}
 
\author{Andrea Cappelli}
\address{I.N.F.N. and Dipartimento di Fisica,
         Largo E. Fermi 2,
         I-50125 Firenze, Italy}
\author{Carlos M\'endez}
\address{Departamento de F\'{\i}sica, Pontif\'{\i}cia Universidade
Cat\'olica, C.P. 38071, 22452-970 Rio de Janeiro,RJ, Brazil} 
\author{Jorge M. Simonin}
\address{Centro At\'omico Bariloche, Comisi\'on Nacional de
         Energ\'{\i}a At\'omica and
         Instituto Balseiro, Universidad Nacional 
         de Cuyo
         8400 -  San Carlos de Bariloche
         R\'{\i}o Negro, Argentina}
\author{Guillermo~R.~Zemba}
\address{Centro At\'omico Bariloche, Comisi\'on Nacional de
         Energ\'{\i}a At\'omica and
         Instituto Balseiro, Universidad Nacional de Cuyo
         8400 -  San Carlos de Bariloche
         R\'{\i}o Negro, Argentina}
 
\maketitle

\begin{abstract}
We present a detailed analysis of the exact numerical spectrum 
of up to ten interacting electrons in the first Landau level
on the disk geometry. 
We study the edge excitations of the hierarchical plateaus and check
the predictions of two relevant conformal field theories:
the multi-component Abelian theory and the $\winf$ 
minimal theory of the incompressible fluids.
We introduce two new criteria for identifying the edge excitations
within the low-lying states: the plot of their density profiles
and the study of their overlaps with the 
Jain wave functions in a meaningful basis.
We find that the exact bulk and edge excitations are very well 
reproduced by the Jain states; these, in turn, can be described by
the multi-component Abelian conformal theory.
Most notably, we observe that the edge excitations form sub-families of 
the low-lying states with a definite pattern, which is explained
by the $\winf$ minimal conformal theory. 
Actually, the two conformal theories are related by
a projection mechanism whose effects are observed in the spectrum.
Therefore, the edge excitations of the 
hierarchical Hall states are consistently described by the
$\winf$ minimal theory, within the finite-size limitations. 
\end{abstract}
 
 


\section{Introduction}

One of the important open problems in the physics of
the quantum Hall effect (QHE) \cite{prange}\cite{dspin}
is the complete understanding of the
hierarchical Hall plateaus, whose filling fractions fall beyond
the Laughlin sequence $\nu=1,1/3,1/5,\dots$ \cite{laugh}.
There are two kinds of theoretical descriptions available at present:
the wave-function constructions and the effective conformal field theories
(CFT) in $(1+1)$ dimensions.

The first approach has culminated in the Jain
theory of the composite-fermion correspondence between
the integer Hall states with $\nu^*=m=2,3,\dots$ and the hierarchical states
with $\nu=m/(mp \pm 1)$, $p=2,4,\dots$, such as 
$\nu=2/5,3/7,\dots$ \cite{jain}.
The existence of the composite-fermion excitations has been 
confirmed by many experiments \cite{cfexp};
the corresponding ansatz wave-functions
have been tested in numerical simulations of few electron systems\footnote{
Moreover, the mean-field theory of the composite fermion has been 
developed in Ref.\cite{mfth}.} \cite{jain}\cite{jj}\cite{jaka}.
These have been mostly done on the spatial geometry of the 
sphere and have firmly established that the Jain states describe
the bulk excitations of quantum Hall fluids. 
On the other hand, these incompressible fluids have 
characteristic edge excitations \cite{wen}, 
which cannot be seen on the sphere geometry. 
These excitations are the relevant low-energy degrees of freedom in the 
conduction experiments, and their basic properties,
like the fractional charge, have been already measured
in the simplest (yet non-trivial) case of the $\nu=1/3$ Laughlin Hall state
\cite{tdom} \cite{mill} \cite{shot}.

The edge excitations are naturally described by 
the conformal field theories\footnote{
An equivalent language is given by the topological Chern-Simons
theories in $(2+1)$ dimensions \cite{juerg}.}
\cite{gins}, because their 
low-energy dynamics is effectively one-dimensional,
being localized on the boundary of the sample  
within a width of the order of the magnetic length $\ell=\hbar c/eB$
\cite{stone}\cite{cdtz1}.
The conformal field theories are a powerful tool because they can be solved 
explicitly in a non-perturbative framework \cite{gins}, and predict 
universal data like the filling fraction, the fractional charge
and quantum statistics of the edge excitations;
moreover, they directly describe the relevant experimental
regime of a large number of electrons.

However, these effective descriptions cannot be easily related
to the microscopic dynamics of the electrons.
Actually, based on simple arguments and symmetry principles,
two classes of CFTs have been proposed for the edge 
excitations of the hierarchical plateaus.
The first is given by the $m$-component Abelian theories $\u1^m$, which 
are generalizations of the successful one-component theory 
for the Laughlin plateaus \cite{read}\cite{juerg}\cite{wen}.
In the literature, these conformal theories have been naturally associated 
with the Jain approach, because the composite-fermion correspondence 
implies the existence of several effective Landau levels, which may
have independent edges of the Laughlin type.

The second class of conformal field theories encompasses the $\winf$ 
minimal models \cite{ctz5}, which exploit the $\winf$ symmetry of the 
incompressible fluids under area-preserving diffeomorphisms of the 
plane \cite{sakita}\cite{ctz1}\cite{ctz3}.
There is a one-to-one correspondence between the hierarchical Hall plateaus
and the $\winf$ minimal theories.
These theories can be obtained by
projecting out some states of the multi-component Abelian theory,
those which are not fluctuations of an elementary incompressible fluid.
This projection implies different properties for 
the edge excitations in the two theories,
which are qualitatively and quantitatively important \cite{ctz5}.

\bigskip

\begin{figure}
\epsfxsize=14cm \centerline{\epsfbox{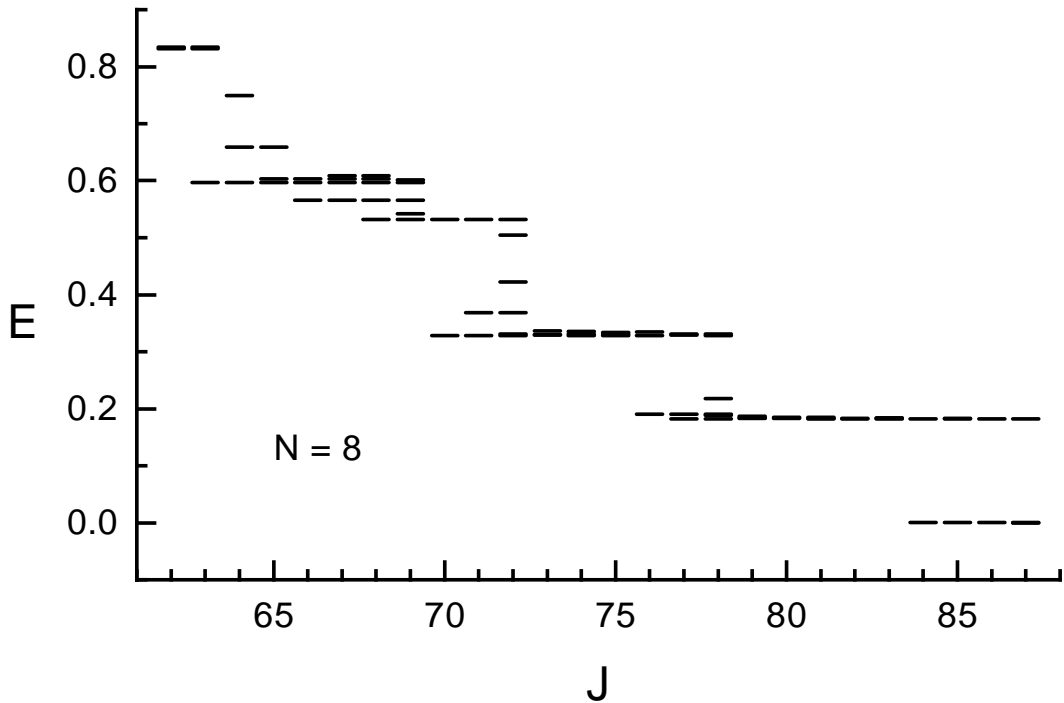}}
\caption{Exact numerical spectrum for $N=8$ on the disk geometry:
only the first few low-lying states are displayed.}
\label{fig1}
\end{figure}

In this paper, we study numerically the spectrum of a finite
systems of $N=6,8$ and $10$ electrons on a disk geometry;
we diagonalize exactly the Hamiltonian with the Haldane short-range 
interaction in the first Landau level \cite{hald}. 
The low-lying states occur in branches which are separated by gaps; 
each branch contains
one state of lowest angular momentum, the ``bottom'' state, 
followed by several higher angular momentum levels with close energies
(see Fig.(\ref{fig1})).
This pattern is well understood for the $\nu=1/3$ Laughlin Hall fluid:
the bottom state is the ground state,
which is an exact eigenstate of the Haldane interaction;
the higher levels are the degenerate edge excitations.
The plot of density profiles shows that the ground state is rather
flat, a characteristic of incompressible fluids \cite{laugh}, and that 
the edge excitations are infinitesimal density deformations.

We use the density plots to gain a qualitative understanding of the 
spectrum: we find that the other bottom states can be either quasi-particle 
excitations (oscillating density profile), or new 
incompressible fluids with lower filling fractions, 
such as $\nu=2/5$ (flat density profile).

Furthermore, we overlap the low-lying states with the Jain ansatz states.
The Jain composite-fermion theory on the disk geometry does predict
the branches in the spectrum: they correspond to
effective Landau levels which are filled selectively at the boundary.
These states are usually denoted by  $[n_1,n_2,\dots,n_k]$, and 
correspond to filling the $i$-th level with $n_i$ electrons, with
$n_1 \ge n_2 \ge \cdots \ge n_k$ and $\sum n_i =N$ \cite{jain}.
Our numerical analysis shows that each branch in the spectrum is
well described by one of these states, corresponding to a given set of 
fillings $\{n_i\}$, and by the corresponding particle-hole excitations.

We then find the following results.
Smooth density profiles identify unambiguously the branches 
corresponding to the $\nu=2/5$ and $ 3/7$ hierarchical plateaus,
for each value of $N$; the rest are quasi-particles branches
over the $\nu=1/3$ Laughlin and the hierarchical plateaus; 
previous numerical analyses \cite{wen}\cite{deja}\cite{kaap}
only charted the energy spectrum and were
naturally led to misinterpret the branches.

Next, we observe that, within each branch, the low-lying states can be 
divided into families of close density profiles: the edge excitations
correspond to infinitesimal deformations of the bottom state,
while families with different density profiles are other bulk 
excitations or magneto-phonons, which are not well separated in
energy for these small values of $N$.
This decomposition in families is first understood in the case
of a quasi-particle over the Laughlin state ($\nu < 1/3$): 
we find a precise correspondence between the exact spectrum 
and the conformal field
theory by using the Jain composite-fermion theory, as it follows.

We begin by computing the overlap matrix between the exact 
low-lying states and the Jain states of the branch (for
example, suppose that the quasi-particle corresponds to the filling 
$[6,2]$ for eight electrons).
These Jain states are particle-hole excitations of two effective Landau 
levels, and can be mapped one-to-one
to the states of the two-component Abelian conformal theory. 
As a consequence, each low-lying exact state in the branch 
has a label of this conformal theory (up to $O(1/N)$ finite-size errors). 
We then find that the edge excitations of the Laughlin quasi-particle
match certain conformal states which are
symmetric with respect to the two Abelian components, i.e. the
two effective Landau levels. 
More precisely, these states are obtained by the projection 
$\u1\times\u1 \to \u1_{\rm diagonal}$ relating the
two-component and the one-component Abelian theories.
Therefore, the Laughlin quasi-particles are indeed
described by the one-component Abelian theory, 
like the well-understood quasi-holes -- 
this is expected in a relativistic theory.
Moreover, the quasi-particle edge excitations amount 
to a specific sub-set of the larger Jain spectrum.

Having understood the pattern of the edge excitations of the 
Laughlin quasi-particles, we proceed to analyse other branches in the
spectrum, whose edge excitations are less understood.
For eight electrons, we find that the edge spectrum of 
the $\nu=2/5$ hierarchical Hall state and its first quasi-particle match the 
predictions of the $\winf$ minimal conformal theory.
This theory realizes the weaker projection
$\u1\times\u1 \to \u1_{\rm diagonal}\times {\rm Virasoro}$ \cite{ctz5}, 
which again eliminates the antisymmetric excitations between the 
two Abelian components, but keeps all the symmetric ones \cite{cz};
there remains a non-trivial sector of neutral excitations,
as explained in Section IIC. 
Therefore, the numerical study support the description of the
hierarchical plateaus by the $\winf$ minimal conformal theory.
The electrons form an irreducible, minimal incompressible fluid,
which is in many respects analogous to the original Laughlin fluid
\cite{laugh}.

The same pattern of families of edge excitations 
in the low-lying spectrum is also found for $N=6$ and $10$ electrons.
For $N=6$, the finite-size span for the angular momentum of edge 
excitations, $\Delta J < O(\sqrt{N})$, is too small to see
any difference between the one-component Abelian (Laughlin states) 
and the $\winf$ minimal theories (hierarchical states);
namely, the results are consistent with our picture but are
not very significative.
For $N=10$, we find that the $\nu=2/5$ Hall plateau clearly displays
the edge spectrum predicted by the $\winf$ minimal theory, as
in the $N=8$ case; on the other hand, the $\nu=3/7$ hierarchical state 
and its quasi-particles are less neat.
Anyhow, these data cannot be consistently interpreted
by the alternative multi-component Abelian theory.

The paper is organized as follows.
In Section $2$, we recall the main results of the Jain theory and
the predictions of the two classes of CFTs, which are then used to
analyse the data in Section $3$ ($N=8$) and Section $4$ ($N=6,10$).
Finally, we end up with some Conclusions.


\section{The Numerical Experiment and its Theoretical Interpretations}

We consider a system of $N=6,8,10$ electrons in the first Landau
level on a disk geometry;
we use open boundary conditions, namely, we truncate the 
Hilbert space of single-particle angular momentum
states at a value that is very large compared to the highest occupied
level in the Laughlin state. 
The interaction among the electrons is given by the Haldane
short-range potential, which selects the $\nu=1/3$ Laughlin
incompressible ground state.
Of course, we are interested in studying the $\nu=2/5$ 
incompressible ground state, and other hierarchical states,
for which there is no analogous model interaction\footnote{
For six electrons, 
we have also computed the spectrum for the Coulomb interaction
and found the same qualitative features.}.

An overview of the $N=8$ spectrum is shown in Fig.(\ref{fig1})
as a function of the angular momentum $J$: 
one clearly sees the branches of low-lying states.
No confining potential is present, but this can be easily added 
afterwards: a quadratic potential increases the energy of each state
by an amount proportional to $J$; thus, 
the bottom state of each branch can be made
the ground state by suitably tuning the strength of the potential.
If this ground state has an approximately constant density, 
its filling fraction is $\nu\sim N(N-1)/2J$, up to finite-size corrections
of relative order $O(1/N)$.
Let us finally quote the previous analysis on the disk geometry,
which were useful in setting up our work \cite{wen}\cite{deja}\cite{kaap}.


\subsection{The Jain Hierarchy}

According to the composite-fermion theory, there is a correspondence
between $m$ filled effective Landau levels 
($\nu^*=m$) and the hierarchical Hall states \cite{jain}:
\beq
\nu^*=m \ \ \ \longleftrightarrow \ \ \ \ \nu={m \over pm \pm 1} \ ,
\qquad m=1,2,\dots, \quad p=2,4,\dots.
\label{cfnu}\eeq
This correspondence is made explicit by the ansatz wave-functions
\beq
\Psi_\nu (z_1,\dots,z_N) =  {\cal P}\ \left(
\prod_{i<j}^N \left(z_i-z_j \right)^2\ \Psi_{\nu^*}(z_1,\dots,z_N)
\right)\ ,
\label{jwf}\eeq
where $\Psi_{\nu^*}$ is the Slater determinant of the filled 
Landau levels and ${\cal P}$ is the projector into the first Landau level.
The composite-fermion correspondence also describes the 
excited states by inserting in (\ref{jwf}) the 
Slater determinants for the $\nu^*=m$ electron transitions.

On the disk geometry, the Landau levels can be filled selectively 
by making electron droplets with different edge shapes: 
these ``bottom'' states are denoted by $[ n_1 , n_2 , \dots, n_k ]$
and correspond to the Slater determinants of $n_i$ 
electrons filling the lowest angular momentum states of
the $i$-th Landau level \cite{deja}.
Each bottom state has an independent branch of low-lying states
corresponding to the particle-hole excitations of the effective levels.
The Jain correspondence (\ref{jwf}) applied to these bottom 
states yields $\nu <1$ states which are characterized by
the following angular momenta and approximate energies \cite{deja}:
\barr
J_{[n_1,\dots,n_k]} &=& N(N-1) + {n_1(n_1-1) \over 2} + {n_2(n_2- 3) \over 2}
+ {n_3(n_3-5) \over 2} + \cdots \ ,\nl
E^{(0)}_{[n_1,\dots,n_k]} &= & E_D \left( n_2 + 2 n_3 + 3 n_4 + 
\cdots \right)\ .
\label{jed}\earr
In the last equation, $E_D$ is the effective Landau level gap,
which has been recently interpreted as the energy for creating
a ``defect'' in the electron fluid\footnote{
The interaction among defects is clearly neglected in $E^{(0)}$.} \cite{jaka}.
In this work, the Jain ansatz was also modified to be 
completely written in the first Landau level;
nevertheless, in this paper we shall use the original proposal
(\ref{jwf}) and perform the projection ${\cal P}$ numerically.

The Jain bottom states (\ref{jed}) for eight electrons
are summarized in Table (\ref{tab1}):
the $[8]$ state is identified as the Laughlin $\nu=1/3$ incompressible
state. The rest of states could be candidates for the $\nu=2/5$
ground state, because they have $J\sim 70$ up to $O(N)$ finite-size
corrections; actually, in the thermodynamic limit $N\to\infty$, 
any state $[N/2 + k, N/2 -k ]$, with $k$ fixed, would work.
The Jain bottom states match rather well the numerical spectrum
in Fig.(\ref{fig1}): 
for almost all predicted $J$ values in Table (\ref{tab1}), there is
a bottom state of a branch of low-lying states;
its energy is rather well approximated by the Jain formula (\ref{jed}). 
On the other hand, a disagreement is seen 
for the smaller $J$ states $[4,4]$ and $[5,2,1]$, which have
$E^{(0)}=4$ but are almost degenerate with the $E^{(0)}=3$ states;
moreover, these two states seem to describe the same branch of levels --
we shall discuss these points later on.


\subsection{Abelian Conformal Field Theories of Edge Excitations}

\begin{figure}
\epsfxsize=14cm \epsfbox{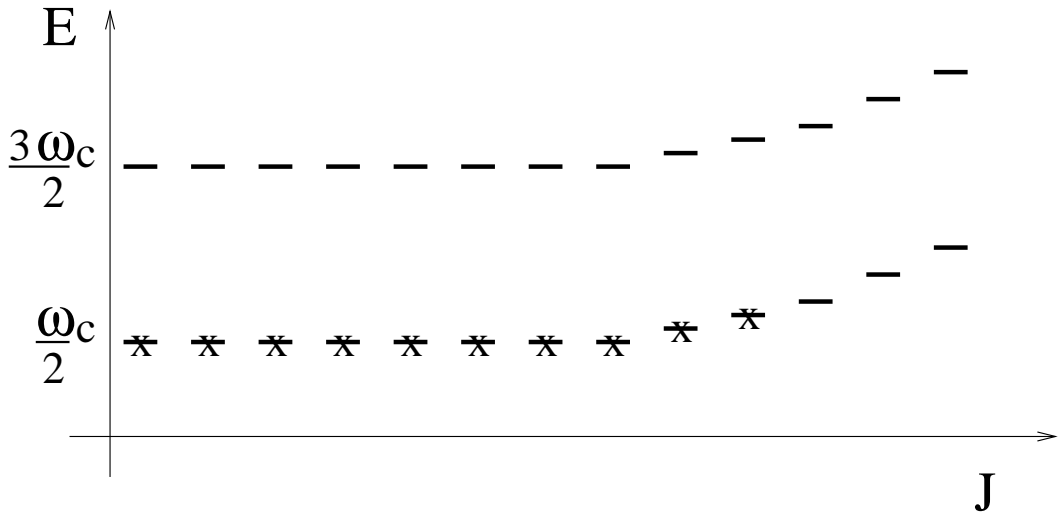} 
\caption{The Landau levels on a disk:
the dashes represent one-particle states, whose energy grows
at large $J$ due to the confining potential.
The crosses represent the electrons filling the first Landau level.}
\label{fig2}
\end{figure}

It is rather well established that the low-energy excitations of a 
droplet of quantum incompressible fluid reside on the
boundary and can be described by a $(1+1)$-dimensional conformal
field theory \cite{wen}.
This can be simply understood in the case of one
filled Landau level (see Fig.(\ref{fig2})), which can be considered
as the Fermi sea (in configuration space) of a one-dimensional system with one 
chirality only \cite{stone}. 
Its low-lying excitations are particle-hole 
transitions near the Fermi surface, which is actually the physical edge of 
the (circular) electron droplet. 
By a well-known procedure, one can take the thermodynamic
limit $N\to\infty$ and approximate the Fermi sea by a Dirac vacuum;
moreover, the energy of the low-lying excitations can be linearized
around the Fermi level, 
$\epsilon\sim v k$, where $k= 2\pi n/R$ is the momentum,
$v$ the Fermi velocity and $R=O(\ell\sqrt{N/\nu})$ the size of the disk.
These are the edge excitations of the $\nu=1$ quantum Hall state;
in a finite system, their wave number is limited by $\vert n\vert \ll R$,
i.e. their angular momentum must be $\Delta J \ll O(\sqrt{N})$
\cite{cdtz1}. Besides the edge excitations,
there are quasi-hole excitations which amount to
moving one electron from deep inside the Fermi sea to the edge
($\Delta J =O(N)$), and, conversely, for the quasi-particles.

In the thermodynamic limit, the relativistic one-dimensional effective
fermion is identified as the charged, chiral Weyl fermion; 
the conformal symmetry of this theory is
described by the Virasoro algebra with central charge $c=1$ \cite{gins}.
This effective conformal field theory of the $\nu=1$ plateau can be 
generalized to the Laughlin 
states by the well-known bosonization procedure:
one rewrites the Weyl fermion in terms of a bosonic field, changes its 
compactification radius and obtains a general $c=1$ 
Abelian conformal theory, whose one-dimensional current
satisfies the Abelian current algebra $\u1$ \cite{cdtz1}; 
an equivalent name is the chiral Luttinger liquid \cite{wen}.
This description of the edge excitations of the Laughlin states
is well established both theoretically and 
experimentally \cite{tdom}\cite{mill}\cite{shot}. 

An important property for the following discussion is that the Hilbert 
spaces of the edge excitations of the integer and Laughlin Hall states,
$\nu=1,1/3,1/5,\dots$, are all isomorphic \cite{wen}; thus, these excitations 
can be still visualized as the particle-hole transitions 
in Fig.(\ref{fig2}).
It is rather simple to count them for small values of $\Delta J=n$, and 
obtain the multiplicities reported in the first line of Table (\ref{tab2})
\cite{wen}; let us stress that the CFT description of edge excitations 
is valid in the linear range $\Delta J < O(\sqrt{N})$ 
\cite{cdtz1}.

A natural generalization of the above picture is to consider the
case of two filled Landau levels: the corresponding edge excitations
are described by the two-component Abelian CFT, which is simply the
tensor product of two one-component theories.
Each Landau level has its own Fermi surface and its particle-hole 
excitations;
their total number is obtained by adding the excitations of the two
levels which have a given $\Delta J=\Delta J_1 +\Delta J_2$
(see the second row of Table (\ref{tab2})).

According to the Jain composite-fermion correspondence (\ref{cfnu}), it is
rather natural to use this two-component Abelian theory for
describing the edge excitations of the hierarchical states (\ref{jwf})
with $\nu=2/(2p+1)=2/5,2/9,\dots$.
As in the one-component case, these $\nu<2$ values can be described
in the conformal theory by bosonizing the two Weyl fermions and by suitably 
changing their compactification radii \cite{wen}.
A characteristic feature of the resulting conformal theories is to realize
an extension of the Abelian current algebra symmetry
$\u1\otimes\u1 \to \u1\otimes\su2$ \cite{read}\cite{juerg}; 
this will be useful in the following discussion.
As in the one-component case, the Hilbert spaces of the edge excitations
above the $\nu=2$ Hall plateau and all the $m=2$ hierarchical states 
(\ref{jwf}) are isomorphic.


\subsection{Minimal Models of Edge Excitations}

Another theory of the hierarchical edge excitations has been
independently proposed in the Ref.\cite{ctz5}: 
it corresponds to the one-component Abelian CFT for the simplest
Laughlin fluids, but differs from the multi-component theory.
It is based on the physical picture of the incompressible 
fluid \cite{laugh}, which possesses the dynamical symmetry under the
area-preserving diffeomorphisms of the spatial coordinates \cite{ctz1}
\cite{sakita}.
This symmetry has been promoted to a building principle for the
CFTs describing its edge excitations \cite{ctz3}: actually, it implies
that the conformal fields should carry a representation of
the $\winf$ algebra, which is a generalization of the 
Abelian current algebra \cite{kac}.

Among the conformal theories with $\winf$ algebra 
\cite{ctz3}, a particular class of models has been found, the 
$\winf$ {\it minimal models} \cite{ctz5}, 
whose filling fractions are in one-to-one 
correspondence with the hierarchical values (\ref{cfnu}).
This is already a strong indication that these models are
experimentally relevant.
The minimal models are characterized by being a reduction
of the previous multi-component Abelian CFTs, in the sense that some
excitations are projected out, as explained hereafter.
This projection implies \cite{ctz5}: (i) a reduced number of edge
excitations above the ground state, as given in the third
line of Table (\ref{tab2}); (ii) only one independent
Abelian charge for the quasi-particles, which is hence identified with
the electric charge; (iii) the existence of neutral quasi-particle 
excitations characterized by
a non-Abelian quantum number of the $SU(m)$ Lie algebra, where $m$ 
is the number of would-be components.

This projection has been recently made explicit
by a Hamiltonian formulation of the minimal incompressible models \cite{cz},
which will be briefly summarized in the first non-trivial case of
two components ($m=c=2$).
Let us start from a closer look into the two-component Abelian theory
with symmetry $\u1\otimes\su2$; for $\nu=2$, this is described by two 
Weyl fermions $\Psi_i(\theta),\overline{\Psi}_i(\theta)$, 
where $\theta$ is the angular variable on the disk and $i=1,2$
denote the upper and lower levels, respectively.
Their excitations can be labelled by the total Abelian charge
$J_0$ and the $SU(2)$ isospin charge $J^3_0$, which are defined as follows:
\barr
J_0 &=& \int_0^{2\pi} {d\theta\over 2\pi} \left( \overline{\Psi}_1\Psi_1
+ \overline{\Psi}_2\Psi_2 \right) \ ,\nl
J_0^3 &=& \int_0^{2\pi} {d\theta\over 2\pi}\ {1\over 2} 
\left( \overline{\Psi}_1\Psi_1 - \overline{\Psi}_2\Psi_2 \right) \ ,\nl
J_0^+ &=& \int_0^{2\pi} {d\theta\over 2\pi}\ \overline{\Psi}_1\Psi_2 \ ,
\qquad\qquad 
J_0^- \ =\ \int_0^{2\pi} {d\theta\over 2\pi}\ \overline{\Psi}_2\Psi_1 \ .
\label{jdef}\earr
One can check that $\left\{J_0^+, J_0^-, J_0^3 \right\}$ satisfy the
$SU(2)$ algebra by using the fermionic canonical commutation 
relations\footnote{
The full set of Fourier modes $\{J^a_n\}$ generate the current
algebra $\su2$ which contains this $SU(2)$ as a subalgebra \cite{gins}.}.
The Abelian and iso-spin charges measure the edge excitations in
the two layers symmetrically and anti-symmetrically, respectively.
Note, however, that the quasi-particles in these Abelian theories
carry the iso-spin quantum number in such a way that it
is linearly additive as another 
electric charge; namely, there are no non-Abelian effects \cite{wen}.
These $SU(2)$ generators can be defined for all the hierarchical 
ground states $\nu=2/5,2/9,\dots$, but are not realized in terms 
of fermions; we can nevertheless continue to use the more intuitive 
fermionic language, owing to the aforementioned isomorphism
between Hilbert spaces.

After these preliminaries, we are ready to define the 
$\winf$ minimal theory (for $c=2$): this is obtained 
from the two-component Abelian theory by imposing 
the constraint \cite{cz},
\beq
J^-_0 \vert {\rm \ minimal \ states\ } \rangle = 0 \ .
\label{hamred}
\eeq
It can be shown that the zero modes 
$\{J_0^\pm, J_0^3 \}$ commute with the Virasoro generators; 
thus, the constraint (\ref{hamred}) does not spoil the conformal
invariance and defines a new conformal theory with the same central 
charge.

The effect of the constraint is the following:
the operator $J^-_0$ moves electrons down and holes up 
between the two layers (with a minus sign in the latter case),
while keeping their (normal-ordered) angular momentum constant;
it relates the edge excitations in the two layers and actually vanishes
on their symmetric linear combinations.
Therefore, the condition (\ref{hamred}) projects out the edge excitations
which are antisymmetric with respect to the two levels.
The ground state is unique and symmetric, then it satisfies 
the constraint: namely, the two CFTs share the same ground state. 

Let us see some examples of allowed edge excitations in the minimal theory.
We first need to clarify the angular-momentum labels in CFT.
The Weyl fermions define a second-quantized relativistic
theory, which describes excitations above the
ground state; therefore, the latter must be specified in order to
relate this description to the numerical data. For example, one
can suppose that the $N=8$ bottom state $[5,3]$ with $J=66$ 
is the $\nu=2/5$ incompressible ground state (see Fig.(\ref{fig1})). 
This fixes the Fermi surface for both layers and sets the reference value for 
the angular momentum of excitations: $\Delta J_1=J_1-4$ and 
$\Delta J_2=J_2-1$. 
The usual moding of conformal fields thus corresponds
to the normal-ordered angular momentum,
which is equal to zero for the charges in (\ref{jdef}). 
The ground state $[5,3]$ and its edge excitations 
are identified by vanishing eigenvalues
of both $J_0$ and $J_0^3$; the other Jain branches in the
spectrum, corresponding to $[6,2]$, etc., in Table(\ref{tab1}),
are described by the CFT as excitations with $J_0^3=-1$, etc,
respectively.

Let us now introduce the fermionic Fock space of the two
Weyl fermions (\ref{jdef}), which also describe the
particle-hole excitations of the two effective Landau levels, 
``up'' and ``down'', in the Jain construction:
the fermionic second-quantized operators are, respectively,
$u_k$, $u^{\dagger}_k$ and $d_k$, $d^{\dagger}_k$  ($k \in {\bf Z}$);
they act on the ground state $\vert \Omega\rangle$. 
There are two Abelian edge excitations at the first excited 
level $\Delta J =1$, which can be written:
\beq
\vert 1; \pm \rangle = \frac{1}{\sqrt{2}} \left( 
d_1^\dagger d_0 \vert \Omega \rangle\ \pm
\ u_1^\dagger u_0 \vert \Omega \rangle \right) .
\label{delta1}\eeq
The constraint (\ref{hamred}) can be written  explicitly: 
$J^-_0 \vert 1; \pm \rangle=
\sum_{k=-\infty}^\infty \ d^\dagger_k u_k\vert 1; \pm \rangle =0$. 
We find that the symmetric combination 
$\vert 1; + \rangle$ satisfies this
constraint and the antisymmetric does not: therefore, the $\winf$
minimal conformal theory only contains the symmetric excitations,
as we anticipated.
At the next level $\Delta J =2$, there are five Abelian edge states:
\barr
\vert\ 2; a \rangle &= &  
d_1^\dagger d_0\ u_1^\dagger u_0 \vert \Omega \rangle\ ,
\nl
\vert\ 2; b\pm \rangle &= &  \frac{1}{\sqrt{2}} \left( 
d_2^\dagger d_0 \vert \Omega \rangle\ \pm
\ u_2^\dagger u_0 \vert \Omega \rangle \right) \ ,
\nl
\vert\ 2; c\pm \rangle &= &  \frac{1}{\sqrt{2}} \left( 
d_1^\dagger d_{-1} \vert \Omega \rangle\ \pm
\ u_1^\dagger u_{-1} \vert \Omega \rangle \right) \ .
\label{deltwo}\earr
The antisymmetric combinations $\vert\ 2; b - \rangle$ and
$ \vert\ 2; c- \rangle$ do not satisfy the constraint (\ref{hamred})
and are not present in the minimal CFT. 
Note that the counting of these states is in agreement with
Table (\ref{tab1}).

In the semiclassical picture developed in Ref.\cite{ctz1}, the 
incompressible Hall fluid is identified with a
Fermi sea and its area-preserving deformations are the 
particle-hole excitations (Fig.(\ref{fig2})). 
The two-level structure introduces 
one additional degree of freedom: the antisymmetric excitations are
tangential to the Fermi surface and do not correspond to 
deformations of the incompressible fluid; therefore, they need not
to be included in a minimal theory. This is the physical meaning
of the condition (\ref{hamred}).

A precise derivation of this constraint can be obtained by analysing  
the irreducible representations of the $\winf$ algebra. 
A general property of CFTs is that they can be constructed by assembling
the representations of their infinite-dimensional symmetry algebra.
The $\winf$ minimal theories were first obtained in such a way
\cite{ctz5}, by using the special, degenerate $\winf$ 
representations, which are equivalent to those of the
algebra $\u1\times {\cal W}_m$, in particular
$\u1\times {\rm Virasoro}$ for $m=2$ \cite{kac}.
The constraint (\ref{hamred}) precisely realizes the 
projection of conformal theories 
$\u1\times\su2 \to \u1\times {\rm Virasoro}$ \cite{cz},
which is a simplified case of the general mechanism of 
Hamiltonian reduction \cite{hred}.

Note that the projection only keeps one state for each $SU(2)$ multiplet
in the Abelian spectrum, i.e. the highest-weight state; 
these states form a non-trivial set of neutral edge excitations, 
which are characterized by their Virasoro weight.
The edge excitations of the other, $\u1_{\rm diagonal}$ part are 
generated by the modes of the symmetric current in (\ref{jdef}),
\beq
J_{-k} =\sum_{j=-\infty}^{\infty} \ 
d^\dagger_{k-j}d_j + u^\dagger_{k-j} u_j \ , \qquad k=1,2,\dots 
\label{udiag}
\eeq
These excitations are symmetric with respect to the two Landau
levels, but are not the most general ones: for example, the state
$\vert 2;a \rangle$ in (\ref{deltwo}) cannot be obtained by applying
(\ref{udiag}) on the ground state.
The symmetric edge excitations obtained by (\ref{udiag}) realize
the reduction of the two-component Abelian theory to the
one-component theory $\u1\times\u1 \to \u1_{\rm diagonal}$,
which also entails a change of central charge $c=2 \to 1$.
 
Finally, we remark that the constraint (\ref{hamred}) can 
be enforced dynamically by modifying
the Hamiltonian of the two-component Abelian theory:
\beq
H \to H +\gamma\ J^+_0 J^-_0 \ \ .
\label{ham}\eeq
The added term is diagonal in the two-component Abelian Hilbert space, 
and is relevant in the renormalization-group sense, because $\gamma$
has dimension of a mass.
It increases the energy of the states which do not satisfy (\ref{hamred})
and in the limit $\ga\to\infty$, it performs the projection leading to
the $\winf$ minimal model.
In general, we may also consider the non-conformal theory with
$\gamma\neq 0,\infty$, which
interpolates between the two-component and the minimal, i.e. 
irreducible, incompressible fluids\footnote{
These are the repulsive and attractive fixed points of the renormalization
group trajectory, respectively.}.
A more complete discussion of these matters can be found in Ref.\cite{cz}.


\section{Analysis of the $N=8$ Data}

We now proceed to analyse each branch of levels in the $N=8$
spectrum and interpret it according to the theories of the previous Section.
The exact eigenstates are denoted by $\Vert\ J - n \rangle\rangle $,
where $J$ is the angular momentum and $n=0,1,2,\dots$ the ordering
by increasing values of the energy. 
The complete set of our numerical data is accessible on-line\footnote{
See: http://andrea.fi.infn.it/cappelli/disk.html.}.

\bigskip

\bigskip

\bigskip

{\bf Branch [8]: the Laughlin incompressible Hall fluid at $\nu=1/3$}

\begin{figure}
\epsfxsize=14cm \centerline{\epsfbox{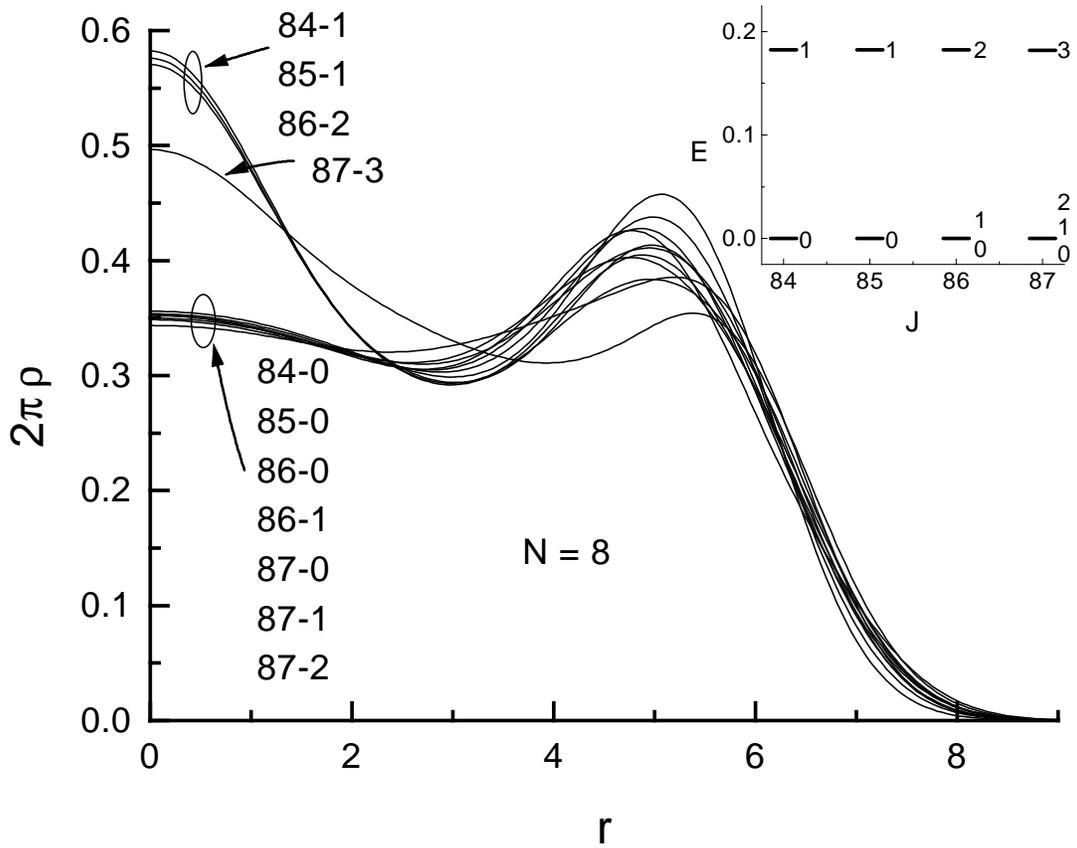}}
\caption{Density profiles and energies for the branch of the $\nu=1/3$ 
Laughlin ground state.}
\label{fig3}
\end{figure}

The density profile $\rho ({\bf x})$ of the Laughlin ground state 
$\Vert 84-0\rr $ is drawn in Fig.(\ref{fig3}): 
we see that this droplet of incompressible  fluid
is fairly flat in the interior; moreover, its value at the origin\footnote{
Hereafter, we set the magnetic length $\ell=1$.}
$2\pi\rho({\bf 0} )$ is close to the 
expected value of $1/3$ in the thermodynamic limit.
The edge excitations are recognized as very small deformations of the
shape of the ground state density; their energies above the ground
state vanish for the Haldane interaction, as shown in 
the inset of Fig.(\ref{fig3}).
Having identified the edge states, we can count their
number: we find the multiplicities $(1,1,2,3,\dots)$ for 
$\Delta J=(0,1,2,3,\dots)$, in agreement with the predictions of the 
one-component Abelian CFT in Table(\ref{tab2}).
The matching of the exact edge excitations with the
the corresponding Jain (Laughlin) wave functions is rather obvious in this
case; the total overlap with each exact numerical state is one by construction.

\bigskip

{\bf Branches [7,1], [6,2]: the quasi-particles over the Laughlin state}

Fig.(\ref{fig30}) shows the density profiles of all the bottom states
found in the $N=8$ spectrum: these are all more oscillating than
the Laughlin ground state. 
The states $\Vert 63-0\rr$ and  $\Vert 70-0\rr$ are clearly quasi-particle
excitations with the characteristic bump of size $O(\ell)$:
therefore, they cannot be interpreted as other incompressible
ground states with $\nu <1/3$.
Let us analyse the $[6,2]$ branch $J=70$ in more detail:
the density profiles and energies of its low-lying states
are drawn in Fig.(\ref{fig4}).
One distinguishes two families of states with opposite oscillations, 
which have multiplicities $(1,1,2)$ and $(0,1,3)$, respectively.  
The first group can be interpreted as a quasi-particle on top of the 
Laughlin $\nu=1/3$ ground state:
actually, the multiplicities of its edge excitations 
agree with the predictions of the one-component Abelian conformal theory
(Table (\ref{tab2})).
The second group can be interpreted as another bulk excitation, 
which might acquire a larger gap in the thermodynamic limit, where
the CFT description of the low-lying states should become exact. 
Although this limit cannot be inferred from the simulation, 
the inset of magnified energy levels in Fig.(\ref{fig4})
clearly shows that these states have a higher energy than those of
the previous group.

Note that the sum of the degeneracies of the two groups 
considered above match the predictions of the two-component Abelian
CFT in Table (\ref{tab2}), i.e. $(1,2,5)$.
By ignoring the information arising from the density plots,
one might be tempted to interpret all these low-lying states
as edge excitations;
then, the corresponding bottom state could not be a Laughlin quasi-particle,
because the edge multiplicities would not match; it would rather be
interpreted as the $\nu=2/5$ incompressible fluid \cite{wen}.
However, the $J=76$ branch, i.e. $[7,1]$, being qualitatively similar,
should be interpreted in the same way, while it is necessarily
a Laughlin quasi-particle, thus arriving to a contradiction.  
We conclude that the new criterion of density plots
is essential for distinguishing Laughlin quasi-particles 
from new hierarchical plateaus.
We can actually observe the formation of a hierarchical
Hall fluid \cite{hiera}: as soon as sizable number $O(N/2)$ of Laughlin
quasi-particles are present, they condense into a new incompressible
fluid, whose density shape is again smooth (see next paragraph).

After having established that $J=70$ is a Laughlin quasi-particle,
we should understand the origin of the extra family $(0,1,3)$
of low-lying states. This is explained by the Jain theory,
which we now analyse in detail;
we are going to illustrate the following chain of relations:
\beq
\begin{array}{ccc}
{low-lying\ exact\ states} 
&\ \  \longleftrightarrow\ \ {Jain\ states}\ \ \longleftrightarrow\ \ 
& { two-component\ Abelian\ CFT} \\
\downarrow & & \downarrow \\
{edge\ excitations} &\longleftrightarrow 
& { one-component\ Abelian\ CFT}
\end{array}
\label{rela}
\eeq
The Jain wave functions corresponding to the $[6,2]$ branch
are obtained by plugging in (\ref{jwf}) the 
Slater determinants for the particle-hole excitations of the first and
second levels filled with $6$ and $2$ electrons, respectively.
These states are, by construction, in one-to-one correspondence with those
of the two-component Abelian conformal theory; thus, we use the
same notation for them, in particular the basis of Section IIC,
equations (\ref{delta1}) and (\ref{deltwo}).
Note, however, that the Jain states are not orthogonal;
therefore, we diagonalize them by the Gram-Schmidt method\footnote{
This introduces some ambiguities for the ordering of states, but they
do not affect the qualitative properties to be discussed hereafter.}
(by also computing the overlaps among themselves).

For each angular momentum value $\Delta J=0,1,2$, we have computed 
the matrix of overlaps between these Jain states and the
low-lying spectrum (see Table (\ref{tab20})). The total square overlap with
each state is very large (about 0.96), and the matrix determinant 
is large enough to match these states one-to-one.
We conclude that the Jain theory describes very well all the 
low-lying states.
Nonetheless, this does not imply that they all should be
edge excitations, as shown by the previous arguments of the density
plot and of the consistency with the Laughlin theory.
Let us recall that, in principle, the Jain theory and the 
conformal field theory give rather different descriptions of the spectrum:
the former applies directly to all excitations for
finite $N$, while the latter only accounts for the edge excitations
in the large $N$ limit, which requires some imagination.

The analysis of the overlap matrices in Table (\ref{tab20}) let us identify
which Jain states correspond to the edge excitations of the Laughlin 
quasi-particle in Fig.(\ref{fig5}).
The edge state $\Vert 71-0\rangle\rangle$ clearly matches the symmetric
state $\vert 1;+\rangle$ in equation (\ref{delta1}). The five states
at $\Delta J=2$ can be divided in two groups: 
($\Vert 72-0\rangle\rangle$, 
$\Vert 72-1\rangle\rangle$, $\Vert 72-3\rangle\rangle$) have
larger overlaps with the symmetric states in (\ref{deltwo}),
and ($\Vert 72-2\rangle\rangle$, $\Vert 72-4\rangle\rangle$) with
the antisymmetric states. 
The precision\footnote{
Table (III) reports approximate numbers for an easy reading;
precise data can be found on-line.} 
is lower than for $\Delta J=1$,
but it is nonetheless remarkable that the signal is not 
completely washed out by the finite-size effects.

This pattern of overlaps can be understood as a relation among 
conformal field theories by applying the analysis of Section IIC:
the edge excitations of the Laughlin quasi-particle,
$\Vert 71-0\rangle\rangle$, 
$\Vert 72-0\rangle\rangle$ and $\Vert 72-1\rangle\rangle$,
have large overlaps with the symmetric states of the two-component Abelian
theory which are generated by the $\u1_{\rm diagonal}$ current
(\ref{udiag}), namely $\vert 1;+\rangle$, $\vert 2;b+\rangle$ and
$\vert 2;c +\rangle$.
Therefore, the Laughlin quasi-particle is indeed described by the
one-component Abelian conformal theory;
the new result is that this is obtained in the Jain spectrum 
by the projection $\u1\times\u1 \to \u1_{\rm diagonal}$
described in the Section IIC.

Similar families of symmetric Laughlin edge excitations with multiplicities 
(1,1,2) are actually found in any branch and for any number of electrons:
this is perhaps the most important result of this paper.
These ``standard'' families are enlarged by further edge
excitations in the case of hierarchical Hall states, as discussed hereafter.

\bigskip

{\bf Branches [5,3], [5,2,1]: the $\nu=2/5$ hierarchical Hall state
and its quasi-particle}

Fig.(\ref{fig5}) shows the family of low-lying states starting at $J=66$, 
which is identified with the $[5,3]$ branch.
The bottom state $\Vert 66-0\rr$ has a rather flat profile
and can be considered as a new incompressible ground
state; the value of $2\pi\rho( {\bf 0} )$ is close to its 
thermodynamic limit of $2/5$ (at variance with the quasi-particle state
$\Vert 76-0\rangle\rangle$).
Moreover, a family of edge excitations is clearly associated to
$\Vert 66-0\rr$, which have multiplicities $(1,1,3)$,
in agreement with the predictions of the $\winf$ minimal CFT 
(third row of Table (\ref{tab2})).

Next, we check whether these edge excitations match the states of the
minimal theory,
by analysing the overlaps with the Jain states (see Table (\ref{tab3})).
We find that the ground state
$\Vert 66-0 \rangle\rangle $ is well-approximated by the Jain bottom 
state $\vert [5,3]\rangle$.
For $\Delta J=1$, the unique edge state $\Vert 67-0 \rr$ is again identified
with the symmetric Jain state $\vert 1; +\rangle$.
At $\Delta J =2$, the exact edge states $\Vert 68- i\rr$, with $i=1,2,3$ 
have large overlaps with the symmetric Jain states 
$\vert\ 2; a \rangle$, $ \vert 2; b +\rangle$ and $\vert 2;c+\rangle$, 
and $O(1/N)$ projection on the antisymmetric Jain states.
Therefore, the edge excitations match all the possible symmetric
Jain states; 
according to Section IIC, these are singled out by the projection 
$\u1\times\u1 \to \u1\times {\rm Virasoro}$, leading to the $\winf$
minimal theory.

Putting all these informations together,
we conclude that this branch can be definitely interpreted as
the $\nu=2/5$ incompressible fluid for $N=8$ electrons, and that its
edge excitations are described by the $\winf$ minimal conformal
theory \cite{ctz5} within the finite-size accuracy.

Let us add one remark. The third state $\Vert 68-0\rr$ at $\Delta J=2$ 
which makes the 
difference between the ubiquitous Laughlin edge spectrum and the
hierarchical one, has an alternative description in the Jain theory
as the bottom state of the $[6,1,1]$ branch.
Here we first encounter the phenomenon of superposition of Jain branches,
which is a sort of degeneracy in the description of the spectrum;
it is an interesting dynamical problem in the Jain theory,
which has not been addressed so far. It can be rephrased as
an ``interaction among defects'' in the regime of compact electron
droplets \cite{jain}; it might also be associated to the mechanism
of the ``condensation of quasi-particles'' of the original hierarchical scheme
\cite{hiera} (or maybe not).
As a matter of fact, this dynamical phenomena does not concern the
CFT interpretation, which is an effective description of the edge 
excitations above a given ground state;
there is just a practical consequence that the one-to-one identification
of states in (\ref{rela}) might be lost in case of mixing of
two competing Jain branches.

The overlaps of the $[6,1,1]$ Jain branch with the
$J=68,69$ low-lying states are not very good, although the bottom
state itself has a good overlap. This branch is rather atypical,
and cannot be considered as a candidate incompressible state
because it is almost gapless with the $[5,3]$ branch.

We now proceed to analyse the branch $J=63$, which is
plotted in Fig.(\ref{fig6}). Its natural interpretation is
of a quasi-particle over the $\nu=2/5$ ground state, 
due to the oscillating density profile. 
The $\winf$ minimal theory predicts the same pattern of edge 
excitations as for the $\nu=2/5$ ground state, i.e. (1,1,3) \cite{ctz5};
this is actually observed in Fig.(\ref{fig6}), and add
further support to whole picture put forward in this paper.
A slight peculiarity is that the third state at $\Delta J=2$ is
higher in energy than other non-edge states -- but the shape is more
important in our analysis.

Next, we proceed to the identification of these states by the
overlap analysis (see Table (\ref{tab40})).
The Jain theory presents a degenerate description for
this branch, namely $[4,4]$ and $[5,2,1]$ (See Table (\ref{tab1})).
The overlaps show that the latter wins, i.e. 
the states of the $[4,4]$ branch have less than $O(1/N)$ projection
on the low-lying states; apparently, the electron tends to 
``pile up'' in the effective Landau levels.
The empirical rule taken from this and other cases of two competing 
Jain branches is that one of them describes well the data,
while the other definitely does not. 

The Jain states of the $[5,2,1]$ branch should be related to a 
three-component Abelian 
conformal theory; this is the first step in extending the chain of
relations (\ref{rela}) to this branch.
For $\Delta J=1$, there are three Abelian edge states, which can be 
written, in analogy with Eq.(\ref{hamred}):
\barr
\vert 1; a \rangle & =& {1\over\sqrt{3}} \left(
u^\dagger_1 u_0 \vert \Omega \rangle +
c^\dagger_1 c_0 \vert \Omega \rangle +
d^\dagger_1 d_0 \vert \Omega \rangle \right) \ ,
\nl
\vert 1; b \rangle & =& {1\over\sqrt{2}} \left(
u^\dagger_1 u_0 \vert \Omega \rangle - c^\dagger_1 c_0 \vert \Omega \rangle 
\right) \ ,\nl
\vert 1; c \rangle & =& {1\over\sqrt{6}} \left(
u^\dagger_1 u_0 \vert \Omega \rangle +
c^\dagger_1 c_0 \vert \Omega \rangle - 2
d^\dagger_1 d_0 \vert \Omega \rangle \right) \ .
\label{three}\earr
The ground state is $\vert\Omega\rangle = \vert [5,2,1]\rangle$ and
the $\{u_k,c_k,d_k\}$ are fermionic Fock operators for
the upper, central and lower (effective) Landau levels.

In this case, we expect that the edge excitations are described
by a reduction from the three-component Abelian theory to the
$c=2$ $\winf$ minimal theory (analogous to the $c=2 \to 1$ projection
(\ref{rela}) for the Laughlin quasi-particles).
It can be shown \cite{cz} that this reduction implies that
the unique edge excitation at $\Delta J=1$ is the completely
symmetric state $\vert 1; a\rangle$. This is in agreement
with the overlaps reported in Table (\ref{tab30}).

Next, the overlap analysis cannot be carried over to 
the $\Delta J=2$ edge excitations, owing to the finite-size limitations:
the Jain state $[5,2,1]$ possesses only one electron in the highest
effective Landau level, and is very far
from the thermodynamic limit of three Fermi surfaces which is implicit in 
the CFT description; in particular, one of the $\Delta J=2$
particle-hole excitations in the higher level is missing.
Therefore, we cannot analyse
the projection of states in this branch; nevertheless, the shape and number
of edge excitations are in agreement with the 
predictions of the $\winf$ minimal theory for a quasi-particle over
the $\nu=2/5$ state.


\section{Analysis of the $N=6$ and $10$ Data}

\subsection{The $N=6$ Data}

Fig. (\ref{fig7}) shows the exact spectrum as a function
of the angular momentum and Table (\ref{tab4}) reports 
the bottom states of the Jain branches with $J\ge 33$.
The observed structure in branches is
similar to that of the $N=8$ case: the Laughlin branch starts 
at $J=45$, and its ground state is identified with the $[6,0]$
Jain state. In decreasing order of $J$, we identify as $[5,1]$
the state appearing at $J=39$ and as $[4,2]$ the one at $J=35$.
For the state appearing at $J=33$ there are two possible Jain
states, $[3,3]$ and $[4,1,1]$, which have the same energy $E^{(0)}$.
Figure (\ref{fig70}) shows the density profiles of all
the bottom states: these give the first hints for identifying
the incompressible ground states. 
The profiles should be compared with the data of average
density $\rho(0)$ and angular momentum for idealized flat
droplets, which are reported in Table (\ref{tab40}).
One finds that the bottom states with $J=39$ and $J=33$
are candidates for the incompressible Hall states with
$\nu=2/5$ and $3/7$, respectively. Let us analyse them in turn.

The profiles of the low-lying states of the $J=39$ branch are
shown in Fig.(\ref{fig71}): the edge states are
$\Vert 40-0 \rr$, $\Vert 41-0\rr$ and $\Vert 41-1 \rr$,
i.e. the observed multiplicities are $(1,1,2)$.
The analysis of the overlaps with the Jain wave functions is similar
to the previous cases: the results are that 
the bottom state $\Vert 39-0\rr$ is identified with $\vert [5,1]\rangle$,
as expected; for $\Delta J=1$, the edge state $ \Vert 40-0\rr$
is the symmetric combination of Abelian edge states
$\vert 1;+\rangle$ (see Eq.(\ref{delta1})); instead, the non-edge 
eigenstate is $\Vert 40-1\rr\sim\vert 1; -\rangle$.
The overlap analysis cannot be extended to $\Delta J=2$, because
the $[5,1]$ ground state has only one electron in the second effective
Landau level, and its particle-hole excitations cannot match the
two-component Abelian CFT.
Within the $\Delta J=1$ analysis, we cannot distinguish
the minimal $c=2$ edge (multiplicities $(1,1,3,\dots)$)
from the $c=1$ Abelian edge $(1,1,2,\dots)$ associated with the
Laughlin quasi-particle excitations.
In conclusion, this branch can be interpreted either as the $\nu=2/5$
incompressible Hall state described by the $\winf$ minimal theory
(but one edge excitation is missing),
or as a Laughlin quasi-particle (but its profile is 
exceptionally flat).

The density profiles of the $J=33$ branch
are presented in Fig. (\ref{fig8}):
the edge excitations are $\Vert 34-0\rangle\rangle$, 
$\Vert 35-1\rangle\rangle$ and $\Vert 35-2\rangle\rangle$, i.e.
again Laughlin-like multiplicities.
Their interpretation in conformal field theory begins by 
identifying the ground state 
$\Vert 33-0\rr$ with one of the two possible Jain bottom states:
the overlaps in Table (\ref{tab5}) shows that the exact state is 
well described by the three-level state $[4,1,1]$, rather than the
``simpler'' $[3,3]$ one; this is the ``piling-up'' of the 
electrons, which we have already encountered.
Actually, the computation of the energies of the two Jain bottom states
shows that the $[3,3]$ branch is separated from the low-lying 
$[4,1,1]$ branch by a gap of order $E_D$; in this case, the interaction 
among defects is of the same order as the energy $E^{(0)}$ 
for non-interacting defects \cite{jain}.

Next, we analyse the low-lying states of this branch: the Jain
excitations of the $[4,1,1]$ bottom state should be compared with
the three-component Abelian CFT, as we have already done for the
branch $[5,2,1]$ of $N=8$ electrons.
Again, we must limit ourselves to the excitations with $\Delta J=1$,
due to the finite sizes of the would-be Fermi seas.
The results for the overlaps in Table (\ref{tab5}) are very
similar to those in Table (\ref{tab30}) for the quasi-particle
over the $\nu=2/5$ state: the unique edge state $\Vert 34-0 \rr$ 
is clearly identified as the symmetric excitation $\vert 1; a\rangle$. 

The  density shape of the $J=33$ bottom state suggests
its interpretation as the $\nu=3/7$ hierarchical Hall state for $N=6$
electrons. Then, we expect that its edge excitations are described by 
the $c=3$ $\winf$ minimal conformal theory, whose multiplicities 
are again $(1,1,3)$ (see the last row of Table (\ref{tab2})) \cite{ctz5};
actually, one can show that the excitations of the $c=3$ and 
$c=2$ minimal theories only differ for $\Delta J \ge 3$, and that 
both agree with those of the simpler Laughlin theory for $\Delta J=1$.

In conclusion, the $N=6$ exact edge states are consistently described
by the $\winf$ minimal conformal theory \cite{ctz5},
within the finite-size limitations: 
we recall that the span for edge excitations in CFT is 
$\Delta J < O(\sqrt{N}) \sim 2$ \cite{cdtz1}.
Another consistent interpretation for all the branches is 
given by the Laughlin one-component theory;
possibly, the condensation of 
quasi-particles leading to the hierarchical Hall fluids
cannot take place in such a small system.
Nevertheless, it is important that the $N=6$ and the $N=8$ data
can be consistently interpreted.


\subsection{\bf The $N=10$ Data}

The spectrum of energies as a function
of the angular momentum for $N=10$ is presented in Fig. (\ref{fig10})
and the branches of Jain states are given in Table (\ref{tab6});
the Laughlin ground state $[10]$ and the $[9,1]$ 
branch are not presented in Fig. (\ref{fig10}), which focusses
on the $\nu\le 2/5$ region. 
The comparison between the exact branches and the Jain predictions 
shows the dynamical phenomenon already seen before: 
for any pair of Jain branches which have degenerate energy
$E^{(0)}$ (Table (\ref{tab6})), only one is realized in the spectrum, 
and correctly describes the low-lying states
(as shown by the overlap analysis).
Actually, $J=111$ matches $[7,3]$, but $[8,1,1]$ is not observed;
similarly $J=108$ matches $[7,2,1]$, $J=103$ is $[6,3,1]$ and
$J=101$ is $[6,2,2]$, while $[6,4]$, $[5,5]$ and $[5,4,1]$ are not observed; 
these are new examples of the ``piling up'' effect.

The analysis of the density profiles of all the bottom
states in Fig.(\ref{fig100})
shows that the branches $J=125$ and $J=117$ are quasi-particles,
with the same qualitative features of the analogous $N=8$ cases.
The bottom states of the $J=111$ and $103$ branches
are not growing in the bulk: although rather oscillating, 
they are the possible candidates
for the $\nu=2/5$ and $\nu=3/7$ incompressible Hall states with $N=10$
electrons, respectively (see the data in Table (\ref{tab40})).

The by-now standard analysis of the edge excitations is first performed
on the $J=111$ branch; the plots are shown in Fig.(\ref{fig101})
and the overlaps with the Jain states are reported in Table (\ref{tab60}).
The usual Laughlin edge excitations ($\Vert 112-0 \rr$,
$\Vert 113-0 \rr$,$\Vert 113-1 \rr$) are clearly seen in the plots
and they definitely overlap on the symmetric Jain states $\vert 1;+\rangle$,
$\vert 2;a\rangle$, $\vert 2;b+\rangle$ and $\vert 2;c+\rangle$. 
According to the $\winf$ minimal conformal theory, the interpretation 
of this branch as the $\nu=2/5$ hierarchical Hall state requires the
identification of a third edge excitation with $\Delta J=2$. 
This can reasonably be $\Vert 113-3\rr$: its density profile is a
non-infinitesimal, but in-phase, deformation of the bottom state, and its
projection is large on the symmetric Jain states (see Table
(\ref{tab60})).
We conclude that the $\winf$ minimal theory consistently describes
this branch as the $\nu=2/5$ hierarchical state:
this is the third definite evidence for this theory. 

The next branch $J=108$, i.e. $[7,2,1]$, is plotted in Fig.(\ref{fig102}):
the profile of the bottom state grows in the bulk 
and this fits the natural expectation that
this branch is a quasi-particle over the $\nu=2/5$ state.
The families of low-lying states are actually
very similar to those of the analogous $N=8$ branch ($J=63$) 
in Fig.(\ref{fig6}).
The multiplicities of the edge excitations should again be (1,1,3) 
as for the $\nu=2/5$ ground state (they are observed in the 
corresponding $N=8$ branch); 
instead the actual computing yields (1,1,2) -- presumably, the missing state
is higher in energy than $\Vert 110-5 \rr$.

Let us now discuss the $J=103$ branch, which is drawn in Fig.(\ref{fig11});
the corresponding overlaps with the Jain branch $[6,3,1]$ 
are reported in Table (\ref{tab7}).
The density profile of the bottom state is similar to
that of the $J=111$ state, i.e. 
the $\nu=2/5$ Hall state; thus, $\Vert 103-0\rr$ can be interpreted as
the $\nu=3/7$ hierarchical state for $N=10$.
The overlap matrix is rather standard for a three-level Jain
branch and identifies the CFT labels of the $\Delta J=1$ low-lying
states; on the other hand, the $\Delta J=2$ excitations 
cannot be analysed for the usual
reason that there is a single electron in the highest effective Landau level. 

The edge excitations are identified by the density profiles and
 their multiplicities are found to be
(1,2,4): this is in disagreement with the 
$c=3$ $\winf$ minimal conformal theory, which predicts the values
(1,1,3) (see Table (\ref{tab2})).
It is possible that there is an accidental degeneracy with another family
(0,1,1-2) of excitations. This interpretation is supported by
the analysis of the next branch $J=101$, which is shown in Fig. (\ref{fig12}).
The bottom state is identified as a quasi-particle over
the $\nu=3/7$ state, as expected; the multiplicities (1,1,\dots) of 
its edge excitations are again in agreement with the $\winf$ minimal theory.

In conclusion, the $N=10$ spectrum presents the general features already
encountered for $N=8$ and $6$; the edge excitations of its hierarchical states 
can be interpreted within
the $\winf$ minimal conformal theory (and the Laughlin quasi-particles
by the one-component Abelian theory, of course).
However, we should remark that the finite-size effects are not
smaller than in the $N=8$ spectrum; this is contrary to the expectation
of a smooth thermodynamic limit towards the conformal field theory.


\section{Conclusions}

In this paper we have presented a comprehensive analysis
of the low-lying spectrum of the electrons in the quantum Hall effect,
in the regime of hierarchical Hall states $\nu\le 2/5$.
We have gone beyond previous works along these
lines: regarding the studies of the composite-fermion correspondence
\cite{deja} \cite{jain}, we have done the first detailed analysis 
of the edge state structure, which was overlooked by the studies on
the spherical geometry \cite{jj}\cite{jaka}. 
We have shown that the Jain composite-fermion theory describes very well
the low-lying spectrum; but we also revealed the dynamical mechanism 
which takes place when two Jain states are allowed:
the electrons tend to ``pile up'', if they are
let to freely fill the effective Landau levels.

Moreover, we improved and reviewed previous analyses on the disk geometry
\cite{wen}\cite{deja}\cite{kaap}. We introduced two new criteria
for the analysis of the exact states: 
i) the plot of their density profiles for distinguishing quasi-particle
excitations from new incompressible ground states, and for identifying the
real edge excitations within the low-lying states;
ii) the interpretation of their overlaps with the Jain states in the 
language of conformal field theory, with concrete relations among
the states and projections thereof.

We have presented a consistent analysis of the low-lying spectrum.
We have shown that the edge excitations form specific subset of 
the low-lying states, by applying the previous criteria and by
checking that the Laughlin quasi-particles are 
described by the well-understood 
one-component Abelian conformal theory.
While all the low-lying states are nicely described by the Jain theory, 
i.e. by the multi-component Abelian conformal theory
\cite{wen}\cite{juerg}\cite{read}, 
the real edge excitations of the hierarchical Hall states 
match the predictions of the $\winf$ minimal theory \cite{ctz5}
(and those of the Laughlin quasi-particles naturally 
match the one-component Abelian theory).
Although the numerical data show some blurs and some finite-size
limitations, the general picture seems firmly established;
we found four neat positive evidences out of the six hierarchical states
with $N=8$ and $10$ electrons.
In conclusion, we hope that this work will stimulate further analyses of
the hierarchical Hall states.

\vspace{2.cm}

{\bf Acknowledgements}

A. C. and G. R. Z. would like to thank the C.E.R.N.
Theory Division and the theory group at L.A.P.P., Annecy, for hospitality.
A. C. also thanks the Theory Group of the Centro At\'omico 
Bariloche for hospitality and acknowledges the partial
support of the European Community Program FMRX-CT96-0012.
G. R. Z. is grateful to I.N.F.N. Sezione di Firenze for hospitality. 
The work of G. R. Z. is supported by a grant of the Antorchas
Foundation (Argentina).

\def\NP{{\it Nucl. Phys.\ }}
\def\PRL{{\it Phys. Rev. Lett.\ }}
\def\PL{{\it Phys. Lett.\ }}
\def\PR{{\it Phys. Rev.\ }}
\def\CMP{{\it Comm. Math. Phys.\ }}
\def\IJMP{{\it Int. J. Mod. Phys.\ }}
\def\MPL{{\it Mod. Phys. Lett.\ }}
\def\RMP{{\it Rev. Mod. Phys.\ }}
\def\AP{{\it Ann. Phys.\ }}



\vspace{2.cm}


\begin{table}
\caption{Bottom states of the Jain theory for $N=8$, ordered by the 
decreasing angular momentum $J \ge 63$, and the corresponding
approximated energies $E^{(0)}$, in units of $E_D$.}
\label{tab1}
\vspace{0.2cm}
\begin{center}
\begin{tabular}{|c|c|c|} 
\hline
$N=8$ Jain states\ \ &  $J$\ \ & $E^{(0)}$\ \ \\ 
\hline
{[8]}     &   84 \   &   0 \ \\
{[7,1]}   &   76 \   &   1 \ \\
{[6,2] }  &   70 \   &   2 \ \\
\hline
{[6,1,1]} &   68  \  &   3 \ \\
{[5,3] }  &   66  \  &   3 \ \\
\hline
{[4,4] }  &   64  \  &   4 \ \\
{[5,2,1]} &   63  \  &   4 \ \\
\hline
\end{tabular}
\end{center}
\end{table}

\begin{table}
\caption{The number of edge excitations for the Laughlin Hall fluid 
$\nu=1/3$ and its quasi-particles (first row); for the hierarchical fluids 
$\nu=2/5$ (second and third rows) and $\nu=3/7$ (forth and fifth rows), 
according to the two relevant conformal field theories with
central charge $c$.}
\label{tab2}
\vspace{0.2cm}
\begin{center}
\begin{tabular}{| c | l | r r r r|}
\hline
$\ \ c\ \ $ & $\Delta J$      & 0 & 1 & 2 & 3 \ \\ 
\hline 
$1$ & One-component Abelian   & 1 & 1 & 2 & 3 \ \\
\hline
$2$ & Two-component Abelian   & 1 & 2 & 5 & 10\ \\
    & Minimal Incompressible  & 1 & 1 & 3 & 5 \ \\
\hline
$3$ & Three-component Abelian & 1 & 3 & 9 & 22\ \\
    & Minimal Incompressible  & 1 & 1 & 3 & 6 \ \\
\hline
\end{tabular}
\end{center}
\end{table}

\begin{table}
\caption{Overlap matrices for a Laughlin quasi-particle branch.
The $N=8$ low-lying exact states are denoted by
$\Vert 70-i \rangle\rangle$, $i=0,1$, 
$\Vert 71-j \rangle\rangle$, $j=0,1$, and 
$\Vert 72-k \rangle\rangle$, $k=0,\dots,4$. 
The corresponding orthogonalized Jain states are
$\vert [6,2]\rangle$, $\vert 1;\pm\rangle$ at $\Delta J=1$, and
($\vert 2;a\rangle,\vert2;b\pm\rangle,\vert 2;c\pm\rangle$) at $\Delta J=2$.} 
\label{tab20}
\vspace{0.2cm}
\begin{center}
\begin{tabular}{|c|c|} 
\hline
J=70\  & $\vert [6,2] \rangle$  \\ 
\hline
$\langle\langle 70-0\Vert$  & 0.985\ \\ 
$\langle\langle 70-1\Vert$  & -0.010\ \\
\end{tabular}
\begin{tabular}{|c|c|c|} 
J=71    & $\vert 1; +\rangle$  & $\vert 1;-\rangle$  \\ 
\hline
$\langle\langle 71-0\Vert$ & -0.98\ & -0.04\  \\
$\langle\langle 72-1\Vert$ &  0.04\ & -0.98\  \\
\end{tabular}
\begin{tabular}{|c|c|c|c|c|c|} 
J=72 & $\vert 2;a\rangle$ & $\vert 2;b+\rangle$ & $\vert 2;b-\rangle$ &
$\vert 2;c+\rangle$ & $\vert 2;c-\rangle$ \\
\hline
$\langle\langle 72-0\Vert$ & 0.63\ & 0.51\ & 0.07\ & 0.45\ & 0.03\ \\
$\langle\langle 72-1\Vert$ & 0.21\ &-0.64\ &-0.37\ & 0.43\ & 0.35\ \\
$\langle\langle 72-2\Vert$ &-0.10\ &-0.05\ & 0.66\ & 0.01\ & 0.74\ \\
$\langle\langle 72-3\Vert$ &-0.66\ & 0.31\ &-0.02\ & 0.72\ &-0.17\ \\
$\langle\langle 72-4\Vert$ & 0.10\ &-0.28\ & 0.52\ & 0.13\ &-0.46\ \\
\hline
\end{tabular}
\end{center}
\end{table}

\begin{table}
\caption{Overlap matrices for the $\nu=2/5$ branch with $N=8$.
The exact states are $\Vert 66-i \rangle\rangle$, $i=0,1$, 
$\Vert 67-j \rangle\rangle$, $j=0,1,2$, and 
$\Vert 68-k \rangle\rangle$, $k=0,\dots,5$;
the orthogonalized Jain states are
$\vert [5,3] \rangle$, $\vert 1;\pm\rangle$ and
($\vert 2;a\rangle,\vert2;b\pm\rangle,\vert 2;c\pm\rangle$).} 
\label{tab3}
\vspace{0.4cm}
\begin{center}
\begin{tabular}{|c|c|} 
\hline
J=66\  & $\vert [5,3] \rangle$  \\ 
\hline
$\langle\langle 66- 0\Vert$  & 0.893\ \\ 
$\langle\langle 66- 1\Vert$  & 0.000\ \\
\end{tabular}
\begin{tabular}{|c|c|c|} 
J=67    & $\vert 1; +\rangle$  & $\vert 1;-\rangle$  \\ 
\hline
$\langle\langle 67-0\Vert$ & 0.890\ & -0.056\  \\
$\langle\langle 67-1\Vert$ & 0.000\ & 0.000 \  \\
$\langle\langle 67-2\Vert$ & 0.001\ & 0.008 \ \\
\end{tabular}
\begin{tabular}{|c|c|c|c|c|c|} 
J=68 & $\vert 2;a\rangle$ & $\vert 2;b+\rangle$ & $\vert 2;b-\rangle$ &
$\vert 2;c+\rangle$ & $\vert 2;c-\rangle$ \\
\hline
$\langle\langle 68-0\Vert$ & -0.283\ & 0.031\  & -0.072\ & 0.351\ & 0.110\ \\
$\langle\langle 68-1\Vert$ &  0.065\ & -0.750\ & -0.145\ & 0.454\ & 0.105\ \\
$\langle\langle 68-2\Vert$ & -0.631\ & -0.366\ & 0.066\  & -0.507\ & 0.051\ \\
$\langle\langle 68-3\Vert$ & 0.000\  & 0.000\  & 0.000\  & 0.000\ & 0.000\ \\
$\langle\langle 68-4\Vert$ & 0.000\  & 0.000\  & 0.000\  & 0.000\ & 0.000\ \\
$\langle\langle 68-5\Vert$ & -0.072\ & -0.155\ & 0.014\  & -0.065\ & 0.128\ \\
\hline
\end{tabular}
\end{center}
\end{table}

\begin{table}
\caption{Overlap matrices of the quasi-particle branch over the
$\nu=2/5$ state for $N=8$. The exact states are
$\Vert 63-i \rangle\rangle$, $i=0,1$, and 
$\Vert 64-j \rangle\rangle$, $j=0,1,2$; 
the orthogonalized Jain states are
$\vert [5,2,1] \rangle$ and its excitations
$\vert 1; x\rangle$, $x=a,b,c$.} 
\label{tab30}
\vspace{0.4cm}
\begin{center}
\begin{tabular}{|c|c|}
\hline 
J=63\  & $\vert [5,2,1] \rangle$\ \\ 
\hline
$\langle\langle 63- 0\Vert$\  &-0.959 \\ 
$\langle\langle 63- 1\Vert$\  & 0.000 \\
\end{tabular}
\begin{tabular}{|c|c|c|c|} 
J=64 & $\vert 1;a\rangle$\ & $\vert 1;b\rangle$\ & $\vert 1;c\rangle$\  \\ 
\hline
$\langle\langle 64-0\Vert$\ & 0.94\ & 0.01\ & 0.14\ \\
$\langle\langle 64-1\Vert$\ & 0.06\ &-0.90\ &-0.32\ \\
$\langle\langle 64-2\Vert$\ &-0.14\ &-0.29\ & 0.87\ \\
\hline
\end{tabular}
\end{center}
\end{table}

\begin{table}
\caption{Jain bottom states for $N=6$, ordered by the 
decreasing angular momentum $J \ge 33$, and the corresponding
energies $E^{(0)}$, in units of $E_D$.}
\label{tab4}
\vspace{0.4cm}
\begin{center}
\begin{tabular}{|c|c|c|} 
\hline
$N=6$ Jain states\ \ &  $J$\ \ & $E^{(0)}$\ \ \\ 
\hline
{[6]}     &   45 \  &   0 \ \\
{[5,1]}   &   39 \  &   1 \ \\
{[4,2] }  &   35 \  &   2 \ \\ 
\hline
{[4,1,1]} &   33 \  &   3 \ \\
{[3,3] }  &   33 \  &   3 \ \\
\hline
\end{tabular}
\end{center}
\end{table}

\begin{table}
\caption{Typical values of the average density $\rho(0)$
and of the angular momentum
for idealized flat droplets of incompressible fluids.}
\label{tab40}
\vspace{0.4cm}
\begin{center}
\begin{tabular}{|c|c|c c c|} 
\hline
$\nu $ &$2\pi\rho(0)$\ \ &         &  $J$ \   &  \\
       &                 & $N=6$ \ & $N=8$ \ & $N=10$ \ \\
\hline
1/3  & 0.33 \ &  45 \  & 84 \ & 135 \ \\
2/5  & 0.40 \ & 37.5 \ & 70 \ & 112.5 \ \\
3/7  & 0.43 \ &  35 \  & 65.3\ & 105 \ \\
\hline 
\end{tabular}
\end{center}
\end{table}

\begin{table}
\caption{Overlap matrices for the candidate $\nu=3/7$ branch with
$N=6$. The exact states $\Vert 33-i \rangle\rangle$, ($i=0,1$), are 
compared with the two Jain bottom states $\vert [3,3] \rangle$ and
$\vert [4,1,1] \rangle$; the $\Delta J=1$ low-lying exact states are
$\Vert 34-j \rangle\rangle$, ($j=0,1,2$), and the Jain states
$\vert 1; x\rangle$, $x=a,b,c$, are excitations
of $\vert [4,1,1] \rangle$.} 
\label{tab5}
\vspace{0.4cm}
\begin{center}
\begin{tabular}{|c|c|c|}
\hline 
J=33\  & $\vert [3,3] \rangle$\ & $\vert [4,1,1] \rangle$\ \\ 
\hline
$\langle\langle 33- 0\Vert$\  &  -0.290\ & 0.935 \\ 
$\langle\langle 33- 1\Vert$\  &\ 0.000\  & 0.000 \\
\end{tabular}
\begin{tabular}{|c|c|c|c|} 
J=34 & $\vert 1;a\rangle$\ & $\vert 1;b\rangle$\ & $\vert 1;c\rangle$\  \\ 
\hline
$\langle\langle 34-0\Vert$\ & -0.872\ & -0.016\ & -0.339\ \\
$\langle\langle 34-1\Vert$\ & -0.091\ & 0.949\  & 0.188\  \\
$\langle\langle 34-2\Vert$\ & 0.000\  & 0.000\  & 0.000\  \\
\hline
\end{tabular}
\end{center}
\end{table}

\begin{table}
\caption{Jain bottom states for $N=10$ with
angular momentum $J \ge 100$, and the corresponding
energies $E^{(0)}$, in units of $E_D$.}
\label{tab6}
\vspace{0.4cm}
\begin{center}
\begin{tabular}{|c|c|c|} 
\hline
$(N=10)$ Jain state\ \ &  $J$\ \ & $E^{(0)}$\ \ \\ 
\hline
{[10]}    &   135 \   &   0 \ \\
{[9,1]}   &   125 \   &   1 \ \\
{[8,2] }  &   117 \   &   2 \ \\
\hline
{[8,1,1]} &   115  \  &   3 \ \\
{[7,3] }  &   111  \  &   3 \ \\
\hline
{[7,2,1]} &   108 \   &   4 \ \\
{[6,4] }  &   107 \   &   4 \ \\
\hline
{[5,5] }  &   105 \   &   5 \ \\
{[6,3,1]} &   103  \  &   5 \ \\
\hline
{[7,1,1,1]} &   105  \  &   6 \ \\
{[6,2,2]}   &   101  \  &   6 \ \\
{[5,4,1]}   &   100  \  &   6 \ \\
\hline
\end{tabular}
\end{center}
\end{table}

\begin{table}
\caption{Overlap matrices for the $\nu=2/5$ branch with $N=10$.
The exact states are $\Vert 111-i \rangle\rangle$, $i=0,1$, 
$\Vert 112-j \rangle\rangle$, $j=0,1$, and 
$\Vert 113-k \rangle\rangle$, $k=0,\dots,4$; 
the orthogonalized Jain states are
$\vert [7,3] \rangle$, $\vert 1;\pm\rangle$ and
($\vert 2;a\rangle,\vert2;b\pm\rangle,\vert 2;c\pm\rangle$).} 
\label{tab60}
\vspace{0.4cm}
\begin{center}
\begin{tabular}{|c|c|} 
\hline
J=111\  & $\vert [7,3] \rangle$  \\ 
\hline
$\langle\langle 111-0\Vert$  &-0.979\ \\ 
$\langle\langle 111-1\Vert$  & 0.014\ \\
\end{tabular}
\begin{tabular}{|c|c|c|} 
J=112    & $\vert 1; +\rangle$  & $\vert 1;-\rangle$  \\ 
\hline
$\langle\langle 112-0\Vert$ & 0.978\ &-0.027\  \\
$\langle\langle 112-1\Vert$ &-0.027\ & 0.940\  \\
\end{tabular}
\begin{tabular}{|c|c|c|c|c|c|} 
J=113 & $\vert 2;a\rangle$ & $\vert 2;b+\rangle$ & $\vert 2;b-\rangle$ &
$\vert 2;c+\rangle$ & $\vert 2;c-\rangle$ \\
\hline
$\langle\langle 113-0\Vert$ & 0.69\ & 0.43\ &-0.03\ & 0.54\ &-0.03\ \\
$\langle\langle 113-1\Vert$ & 0.04\ &-0.77\ &-0.17\ & 0.56\ & 0.14\ \\
$\langle\langle 113-2\Vert$ &-0.06\ & 0.00\ &-0.59\ & 0.00\ &-0.73\ \\
$\langle\langle 113-3\Vert$ &-0.49\ & 0.16\ & 0.46\ & 0.50\ &-0.33\ \\
$\langle\langle 113-4\Vert$ &-0.20\ & 0.13\ &-0.14\ & 0.15\ & 0.13\ \\
\hline
\end{tabular}
\end{center}
\end{table}

\begin{table}
\caption{Overlap matrices for the $\nu=3/7$ branch for $N=10$.
The exact states are $\Vert 103-i \rr$, $i=0,1$, and 
$\Vert 104-j \rangle\rangle$, $j=0,1,2$; 
the orthogonalized Jain states are
$\vert [6,3,1] \rangle$ and its excitations
$\vert 1; x\rangle$, $x=a,b,c$.} 
\label{tab7}
\vspace{0.4cm}
\begin{center}
\begin{tabular}{|c|c|} 
\hline
J=103\ & $\vert [6,3,1] \rangle$  \\ 
\hline
$\langle\langle 103- 0\Vert$ & -0.954 \\ 
$\langle\langle 103- 1\Vert$ &  0.000 \\ 
\end{tabular}
\begin{tabular}{|c|c|c|c|} 
J=104 & $\vert 1;a\rangle$ & $\vert 1;b\rangle$ & $\vert 1;c\rangle$  \\ 
\hline
$\langle\langle 104-0\Vert$\ & -0.947\ & 0.030\ & -0.105\ \\
$\langle\langle 104-1\Vert$\ &  0.031\ &-0.779\ & -0.507\ \\
$\langle\langle 104-2\Vert$\ &  0.092\ & 0.376\ & -0.715\ \\
\hline
\end{tabular}
\end{center}
\end{table}


\begin{figure}
\epsfxsize=14cm \centerline{\epsfbox{8r.eps}}
\caption{Comparison of the density profiles for all
the $N=8$ bottom states.}
\label{fig30}
\end{figure}

\begin{figure}
\epsfxsize=14cm \centerline{\epsfbox{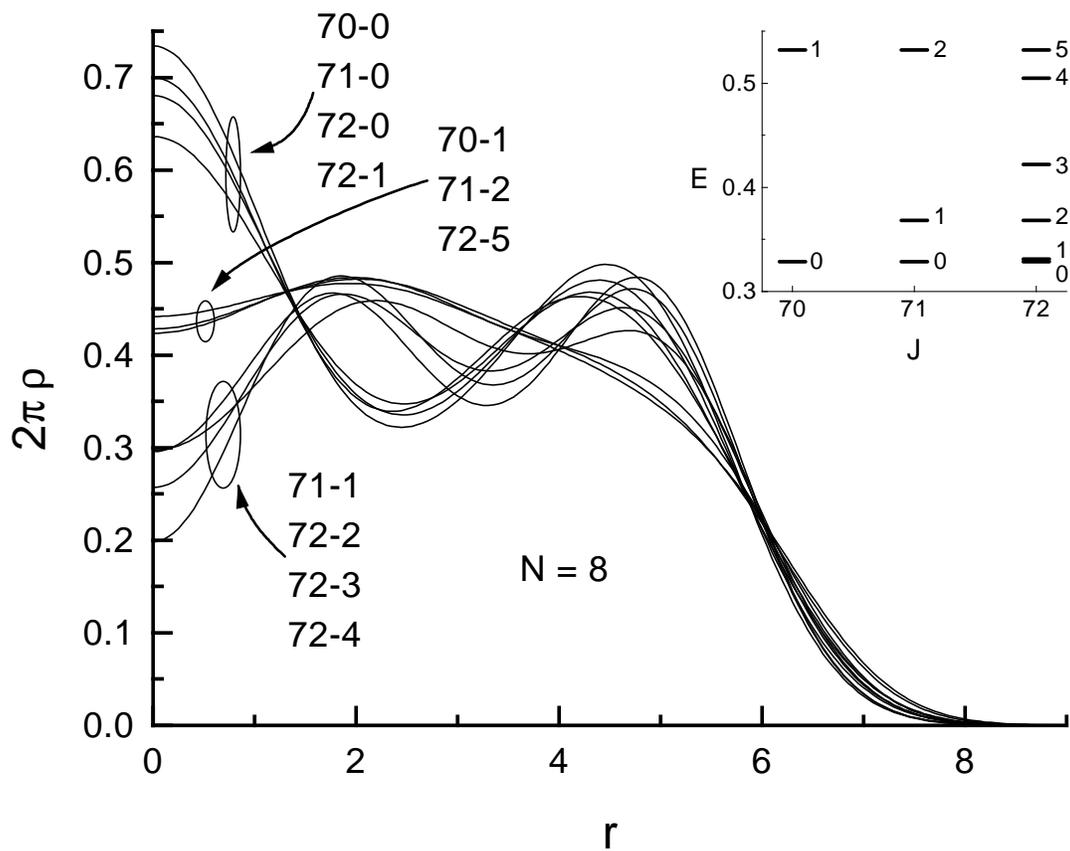}}
\caption{Density profiles and energies of a quasi-particle branch
over the Laughlin state.}
\label{fig4}
\end{figure}

\begin{figure}
\epsfxsize=14cm \centerline{\epsfbox{8-66.eps}}
\caption{Density profiles and energies for the branch of the 
$\nu=2/5$ hierarchical ground state.}
\label{fig5}
\end{figure}

\begin{figure}
\epsfxsize=14cm \centerline{\epsfbox{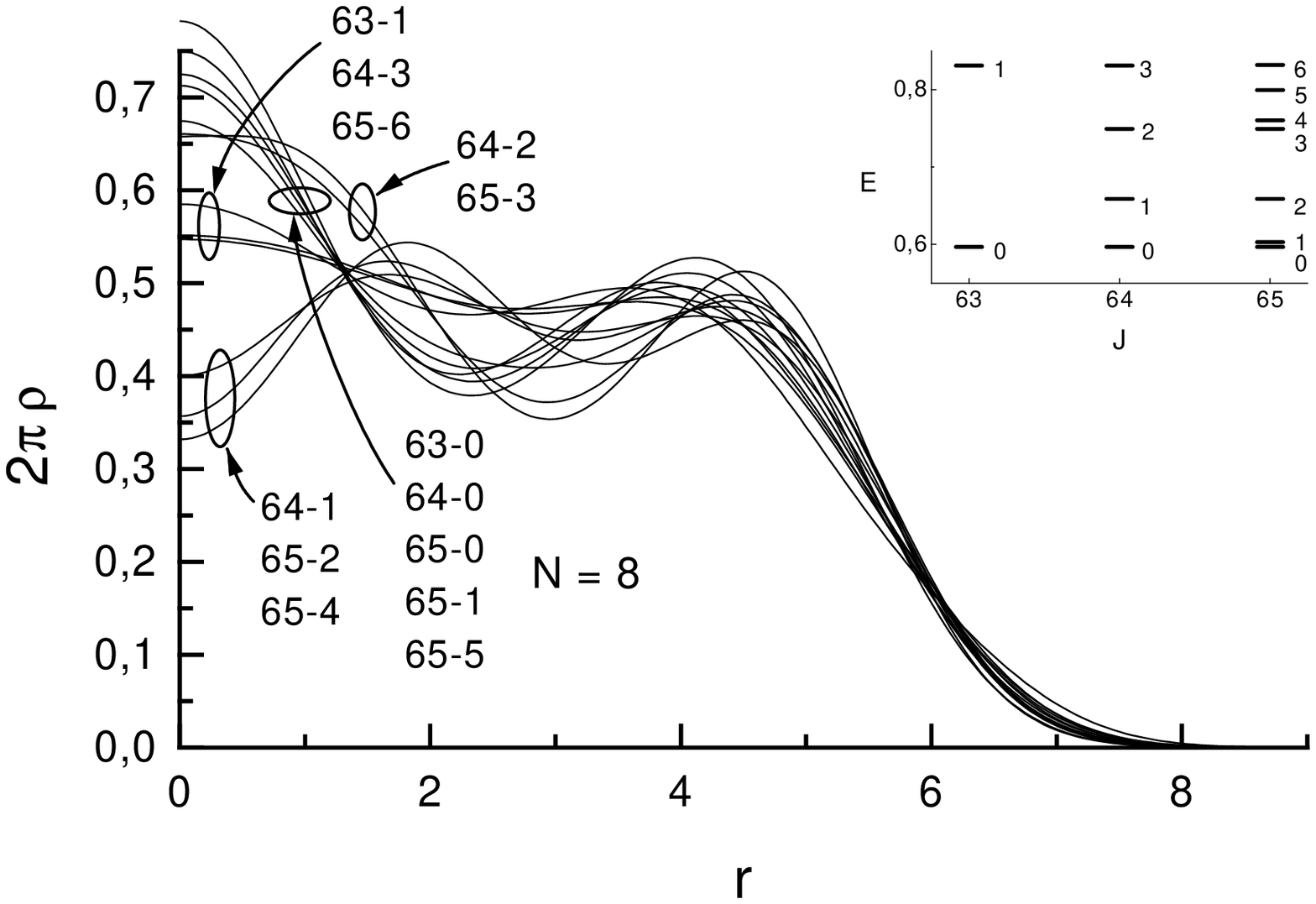}}
\caption{Density profiles and energies for the quasi-particle branch 
over the $\nu=2/5$ state.}
\label{fig6}
\end{figure}

\begin{figure}
\epsfxsize=14cm \centerline{\epsfbox{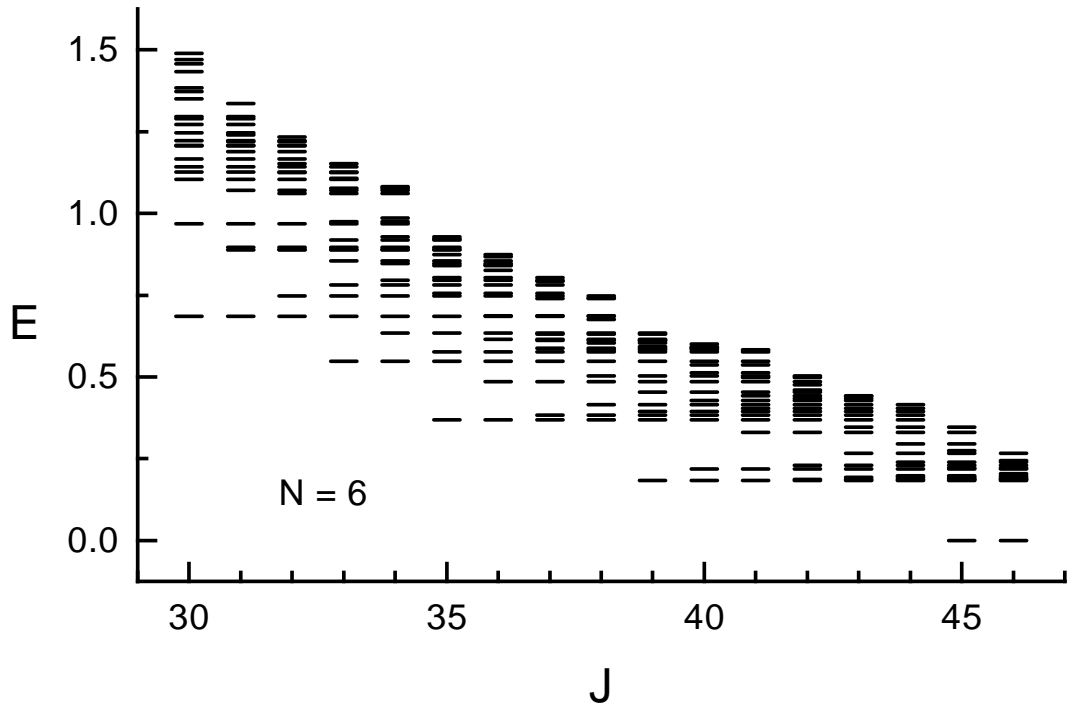}}
\caption{Exact numerical spectrum for $N=6$.}
\label{fig7}
\end{figure}

\begin{figure}
\epsfxsize=14cm \centerline{\epsfbox{6r.eps}}
\caption{Comparison of the density profiles for all
the $N=6$ bottom states.}
\label{fig70}
\end{figure}

\begin{figure}
\epsfxsize=14cm \centerline{\epsfbox{6-39.eps}}
\caption{Density profiles of $N=6$ branch which can be interpreted 
as the $\nu=2/5$ ground state.}
\label{fig71}
\end{figure}

\begin{figure}
\epsfxsize=14cm \centerline{\epsfbox{6-33.eps}}
\caption{Density profiles of $N=6$ branch which can be interpreted 
as the $\nu=3/7$ ground state.}
\label{fig8}
\end{figure}

\begin{figure}
\epsfxsize=14cm \centerline{\epsfbox{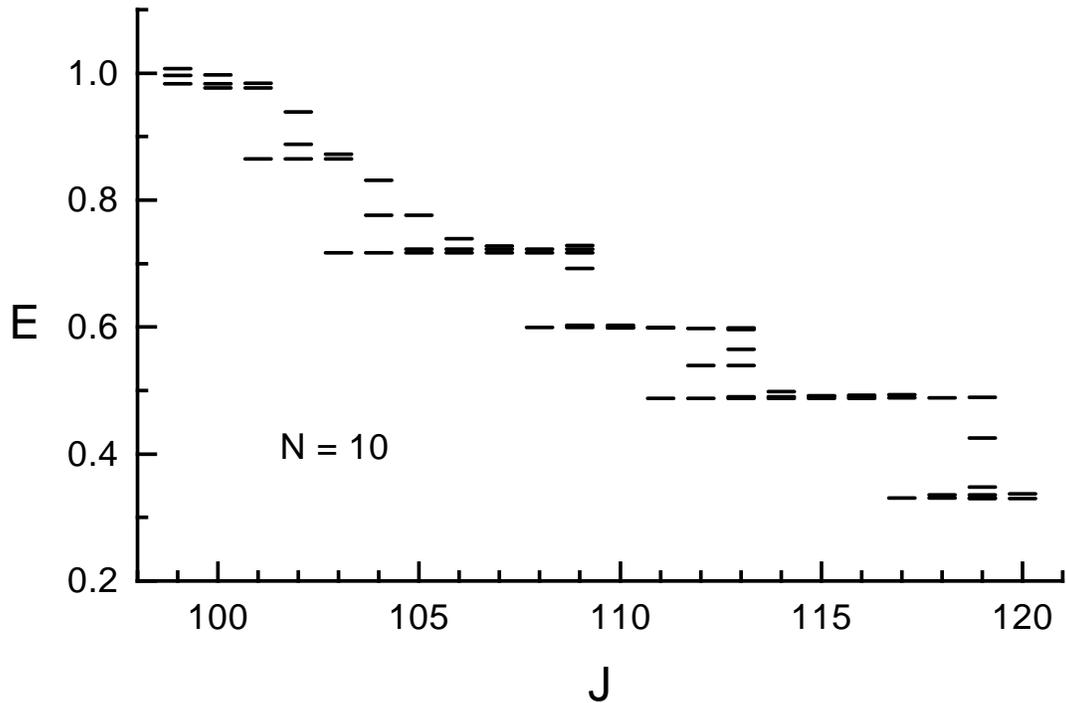}}
\caption{Exact numerical spectrum for $N=10$:
only the  first few low-lying states are displayed.}
\label{fig10}
\end{figure}

\begin{figure}
\epsfxsize=14cm \centerline{\epsfbox{10r.eps}}
\caption{Comparison of the density profiles for all
the $N=10$ bottom states.}
\label{fig100}
\end{figure}

\begin{figure}
\epsfxsize=14cm \centerline{\epsfbox{10-111.eps}}
\caption{Density profiles and energies for the branch 
of the $\nu=2/5$ ground state.}
\label{fig101}
\end{figure}

\begin{figure}
\epsfxsize=14cm \centerline{\epsfbox{10-108.eps}}
\caption{Density profiles and energies for the branch 
of a quasi-particle over the $\nu=2/5$ state.}
\label{fig102}
\end{figure}

\begin{figure}
\epsfxsize=14cm \centerline{\epsfbox{10-103.eps}}
\caption{Density profiles and energies for the branch 
of the $\nu=3/7$ ground state.}
\label{fig11}
\end{figure}

\begin{figure}
\epsfxsize=14cm \centerline{\epsfbox{10-101.eps}}
\caption{Density profiles and energies for the branch 
of a quasi-particle over the $\nu=3/7$ state.}
\label{fig12}
\end{figure}


\begin{references}
\bibitem{prange} For a review see: R. A. Prange, S. M. Girvin, 
                {\it The Quantum Hall Effect}, Springer Verlag, 
                New York (1990).
\bibitem{dspin} For a review see: S. Das Sarma and A. Pinczuk,
                {\it Perspectives in Quantum Hall Effects}, Wiley,
                New York (1996).
\bibitem{laugh} R. B. Laughlin, \PRL {\bf 50} (1983) 1395;
                for a review see: R. B. Laughlin, {\it Elementary Theory: the
                Incompressible Quantum Fluid}, in \cite{prange}.
\bibitem{jain}  J. K. Jain, \PRL {\bf 63} (1989) 199; \PR {\bf B 41} (1990)
                7653; for reviews see: J. K. Jain, {\it Adv. in Phys.} 
                {\bf 41} (1992) 105, and 
                {\it Composite Fermions}, in \cite{dspin}. 
\bibitem{cfexp} For a review see: H. L. Stormer and D. C. Tsui, 
                {\it Composite Fermions in
                the Fractional Quantum Hall Effect}, in \cite{dspin}
\bibitem{mfth}  For a review see: E. Fradkin and A. Lopez, 
                \NP {\bf B (Proc. Suppl.) 33C} (1993) 67.
\bibitem{jj}    X. G. Wu and J. K. Jain, \PR {\bf B 51} (1995) 1752.
\bibitem{jaka}  J. K. Jain and R. K. Kamilla, \IJMP {\bf B 11} (1997) 2621.
\bibitem{wen}   For a review, see: X. G. Wen, \IJMP {\bf 6 B} (1992) 1711,
                {\it Adv. in Phys.} {\bf 44} (1995) 405.
\bibitem{tdom}  R. C. Ashoori, H. L. Stormer, L. N. Pfeiffer, K. W. Baldwin
                and K. West, \PR {\bf B 45} (1992) 3894;
\bibitem{mill}  F. P. Milliken, C. P. Umbach and R. A. Webb, 
                {\it Solid State Commun.} {\bf 97} (1996) 309;
                P. Fendley, A. W. W. Ludwig and
                H. Saleur, \PR {\bf B 52} (1995) 8934;
                for a review, see: C. L. Kane and M. P. A. Fisher,
                {\it Edge-State Transport}, in \cite{dspin}.
\bibitem{shot}  V. J. Goldman and B. Su, {\it Science} {\bf 267} (1995)
                1010; R. de-Picciotto et al., cond-mat/9707289;
                L. Saminadayar et al., cond-mat/9706307.
\bibitem{juerg} J. Fr\"ohlich and A. Zee, \NP {\bf 364 B} (1991) 517; 
                X.-G. Wen and A. Zee, \PR {\bf 46 B} (1993) 2290.
                J. Fr\"ohlich and E. Thiran, {\it J. Stat. Phys.}
                {\bf 76} (1994) 209; J. Fr\"olich, T. Kerler, U. M. Studer 
                and E. Thiran, \NP {\bf B 453} (1995) 670.
\bibitem{gins}  A. A. Belavin, A. M. Polyakov and A. B. Zamolodchikov,
                \NP {\bf B 241} (1984) 333; for a review see:
                P. Ginsparg, {\it Applied Conformal Field Theory},
                in {\it Fields, Strings and Critical Phenomena},
                Les Houches School 1988, E. Brezin and J. Zinn-Justin eds.,
                North-Holland, Amsterdam (1990).
\bibitem{stone} M. Stone, \AP (NY) {\bf 207} (1991) 38. 
\bibitem{cdtz1} A. Cappelli, G. V. Dunne, C. A. Trugenberger and G. R.
                Zemba, \NP {\bf 398 B} (1993) 531.
\bibitem{read}  N. Read, \PRL {\bf 65} (1990) 1502.
\bibitem{ctz5}  A. Cappelli, C. A. Trugenberger and G. R. Zemba,
                \NP {\bf 448} (1995) 470; for a review, see: 
                \NP {\it (Proc. Suppl.)} {\bf B 45A} (1996) 112.
\bibitem{sakita}S. Iso, D. Karabali and B. Sakita,
                \NP {\bf B 388} (1992) 700, \PL {\bf B 296} (1992) 143.
\bibitem{ctz1}  A. Cappelli, C. A. Trugenberger and G. R. Zemba,
                \NP {\bf 396 B} (1993) 465, \PL {\bf 306 B} (1993) 100;
                for a review, see:
                A.Cappelli, G.V.Dunne, C.A.Trugenberger and G.R.Zemba,
                \NP {\bf B (Proc. Suppl.) 33C} (1993) 21.
\bibitem{ctz3}  A. Cappelli, C. A. Trugenberger and G. R. Zemba,
                \PRL {\bf 72} (1994) 1902.
\bibitem{hald}  F. D. M. Haldane, {\it The Hierarchy of Fractional States and
                Numerical Studies}, in \cite{prange}.
\bibitem{deja} G. Dev and J. K. Jain, {\it Phys. Rev.} {\bf 45 B} (1992) 1223.
\bibitem{kaap} M. Kasner and W. Apel, {\it Phys. Rev.} {\bf 48 B}
               (1993) 11435; {\it Ann. Physik} {\bf 3} (1994) 433.             
\bibitem{cz}   A. Cappelli and G. R. Zemba, 
                {\it Hamiltonian Formulation for the Minimal Models
                of the Incompressible Quantum Hall Fluids}, to appear.
\bibitem{kac}   V. Kac and A. Radul, {\it Comm. Math. Phys.} {\bf 157}
                (1993) 429; H. Awata, M. Fukuma, Y. Matsuo and
                S. Odake, {\it Prog. Theor. Phys. (Supp.)} {\bf 118}
                (1995) 343; E. Frenkel, V. Kac, A. Radul and W. Wang,
                \CMP {\bf 170} (1995) 337.
\bibitem{hred} A. M. Polyakov, \IJMP {\bf A 5} (1990) 833; 
               M. Bershadsky and H. Ooguri, \CMP {\bf 126} (1989) 49. 
\bibitem{hiera}F. D. M. Haldane \PRL {\bf 51} (1983) 605;
               B. I. Halperin, \PRL {\bf 52} 91984) 1583;
               M. Greiter, \PL {\bf B 336} (1994) 48.
\end{references}
\end{document}